\documentclass{aa}  
\usepackage{graphicx}
\usepackage{txfonts}
\usepackage{natbib}
\usepackage{multirow}

\begin{document}

\title{GRB 100219A with X-shooter - abundances in a galaxy at $z=4.7$ }

\author{C. C. Th\"one \inst{1,2}, J. P. U. Fynbo \inst{3}, P. Goldoni \inst{4}, A. de Ugarte Postigo \inst{1,3},  S. Campana \inst{5},  S. D. Vergani \inst{6},  S. Covino \inst{5}, T. Kr\"uhler \inst{3,7,8}, L. Kaper \inst{9}, N. Tanvir \inst{10}, T. Zafar \inst{3}, V. D'Elia \inst{11,12}, J. Gorosabel \inst{1}, J. Greiner \inst{7}, P. Groot \inst{13}, F. Hammer \inst{6}, P. Jakobsson \inst{14}, S. Klose \inst{15}, A. J. Levan \inst{16}, B. Milvang-Jensen \inst{3}, A. Nicuesa Guelbenzu \inst{15}, E. Palazzi \inst{17}, S. Piranomonte \inst{11}, G. Tagliaferri \inst{5},  D. Watson \inst{3}, K. Wiersema \inst{10}, R. A. M. J. Wijers \inst{9}}

\offprints{C.C. Th\"one}

\institute{Instituto de Astrof\'isica de Andaluc\'ia, CSIC, Glorieta de la Astronom\'ia s/n, E - 18008 Granada, Spain  \email{cthoene@iaa.es}
\and Niels Bohr International Academy, Niels Bohr Institute, Blegdamsvej 17, 2100 Copenhagen, Denmark
\and Dark Cosmology Centre, Niels Bohr Institute, University of Copenhagen, Juliane Maries Vej 30, 2100 K\o benhavn \O, Denmark
\and APC,AstroParticule et Cosmologie, UniversitŽ Paris Diderot, CNRS/IN2P3, CEA/Irfu, Observatoire de Paris, Sorbonne Paris CitŽ, 10 rue Alice Domon et Leonie Duquet, 75205 Paris Cedex 13, France
\and INAF, Osservatorio Astronomico di Brera, Via E. Bianchi 46, 23807 Merate, Italy
\and GEPI, Observatoire de Paris, CNRS, Univ. Paris Diderot, 5 place Jules Janssen, 92190 Meudon, France
\and Max-Planck-Institut f\"{u}r extraterrestrische Physik, Giessenbachstra\ss e, 85748 Garching, Germany
\and Excellence Cluster Universe, Technische Universit\"{a}t M\"{u}nchen, Boltzmannstra\ss e 2, 85748, Garching, Germany.
\and Astronomical Institute Anton Pannekoek, University of Amsterdam, Science Park 904, 1098 XH Amsterdam, the Netherlands
\and Department of Physics and Astronomy, University of Leicester, University Road, Leicester LE1 7RH, UK
\and INAF, Osservatorio Astronomico di Roma, via di Frascati 33, 00040 Monte Porzio Catone, Rome, Italy
\and ASI-Science Data Center, via Galileo Galilei, 00044 Frascati, Italy 
\and Department of Astrophysics, IMAPP, Radboud University Nijmegen, PO Box 9010, 6500 GL Nijmegen, the Netherlands
\and Centre for Astrophysics and Cosmology, Science Institute, University of Iceland, Dunhagi 5, 107 Reykjav\'ik, Iceland
\and Th\"uringer Landessternwarte Tautenburg, Sternwarte 5, 07778 Tautenburg, Germany
\and Department of Physics, University of Warwick, Coventry, CV4 7AL, UK
\and INAF, IASF di Bologna, via Gobetti 101, 40129 Bologna, Italy
}

      \date{Received; accepted }

\begin{abstract}
{Abundances of galaxies at redshifts $z>4$ are difficult to obtain from damped Ly $\alpha$ (DLA) systems in the sightlines of quasars (QSOs) due to the Ly $\alpha$ forest blanketing and the low number of high-redshift quasars detected. Gamma-ray bursts (GRBs) with their higher luminosity are well suited to study galaxies out to the formation of the first stars at $z>10$. Its large wavelength coverage makes the X-shooter spectrograph an excellent tool to study the interstellar medium (ISM) of high redshift galaxies, in particular if the redshift is not known beforehand.}
{We want to determine the properties of a GRB host at $z=4.66723$ from absorption lines. This is one of the highest redshifts where a detailed analysis with medium-resolution data is possible. Furthermore, we determine the dust extinction using the spectral energy distribution from X-rays to near infrared. Finally, we put the results in context to other GRBs at $z>4$ and properties of (high redshift) QSO absorbers.}
{The velocity components of the resonant and fine-structure absorption lines are fitted with Voigt-profiles and the metallicity determined from S, Si, Fe and O. \ion{Si}{II*} together with the energy released in the UV restframe as determined from GROND photometric data gives us the distance of the absorbing material from the GRB. The extinction is determined from the spectral slope using X-ray spectral information and the flux calibrated X-shooter spectrum, cross calibrated with photometric data from GROND. We also collect information on all GRB hosts with $z>4$.}
{We measure a relatively high metallicity of $\textnormal{[M/H]}=-1.0\pm0.1$ from S, the distance of the material showing fine-structure lines is  0.3 to 1.0\,kpc. The extinction is moderate with $\textnormal{A}_V=0.24\pm0.06$\,mag. Low- and high ionization as well as fine-structure lines show a complicated kinematic structure probably pointing to a merger in progress. We detect one intervening system at $z=2.18$. }
{GRB-DLAs have a shallower evolution of metallicity with redshift than QSO absorbers and no evolution in their HI column density or ionization fraction. GRB hosts at high redshift seem to continue the trend of the metallicity-luminosity relation towards lower metallicities but the sample is still too small to draw a definite conclusion. While the detection of GRBs at z $>$ 4 with current satellites is still difficult, they are very important for our understanding of the early epochs of star- and galaxy-formation in the Universe.}
\end{abstract}


\authorrunning{C. C. Th\"one et al.}
\titlerunning{GRB 100219A with X-shooter}
\maketitle

\section{Introduction}
Determining the properties of the interstellar medium (ISM) in high redshift
galaxies is difficult due to their extreme faintness. Detailed and resolved studies using emission lines in the optical/near-IR range is
restricted to redshifts up to $z\sim3$ with current facilities (see, e.g.,
\citealt{FoersterSchreiber, Maiolino08}) and only with the help of gravitational lensing beyond this redshift (see, e.g., \citealt{Swinbank07} for resolved studies of a gravitationally lensed galaxy at $z=4.9$). The use of GRBs as lighthouses, similar to what has
been done since many years with intervening absorption systems towards QSOs, is a 
promising way to study the properties of galaxies $z > 4$. In contrast to QSOs, GRBs with their higher luminosities can probe more central and denser regions of those galaxies. Long GRBs are connected to massive star-formation (see, e.g., \citealt{WoosleyBloom} for a review of the GRB-supernova connection), hence GRBs probe star-forming places across cosmic history (e.g., \citealt{Fruchter}).

GRB afterglows constitute a powerful source for absorption line studies as they have smooth, featureless intrinsic spectra on
which the intervening ISM in the host imprints absorption lines from different
metals. Optical spectroscopy allows the determination of metallicities from $z\sim1.7$ to $\sim6$ when those metal absorption lines get redshifted into the near-infrared (NIR).
X-shooter at the ESO Very Large Telescope (VLT) \citep{DOdorico, Vernet11} with its wavelength coverage from 3,000 to 24,800 {\AA}  facilitates detailed investigations of galaxies up to redshifts of $z\sim11-13$, provided GRBs exist at such high redshifts. The broad wavelength coverage of X-shooter also
allows to fit an extinction curve and to derive the reddening from the shape of the continuum, assuming an intrinsic powerlaw for the continuum. In most cases, its medium resolution is also good enough to allow the fitting of the absorption lines with Voigt profiles, allowing a more accurate determination of the column density compared to low resolution methods. 

Previous studies have revealed a diverse picture of GRB host metallicities but they suffer from a small sample size, with currently about 20 metallicities published in the literature (for the largest samples see, e.g., \citealt{Fynbo06, Prochaska07,Savaglio09}). Metallicities of GRB hosts seem to be on average higher than those obtained from QSO-DLAs, in particular at higher redshifts (e.g. \citealt{Fynbo06}). GRB host metallicities also seem to show a slower evolution with redshift, albeit with a large scatter at any given redshift. Both GRB and QSO samples, however, show subsolar metallicities, in some cases down to 1/100 solar or less (e.g., GRB 050730, \citealt{DElia07}, and GRB 090926A, \citealt{Rau090926A, DElia10}). The difference between the two samples can be explained by QSO absorbers probing the outer, less metal-rich regions of galaxies due to their lower luminosity and the sight-lines being selected according to the cross-section, and by GRB hosts being larger than QSO-DLA galaxies \citep{Fynbo08}.

The gas observed in GRB absorption-line studies is mostly in a low ionization state (see, e.g., \citealt{Fynbo09}) which makes it a good representation of the cold gas in the host galaxy, mostly unaffected by the GRB itself. The same is the case for the material observed in QSO absorbers; however, one has to consider that QSO and GRB-DLAs are likely probing different regions in the galaxy. Exceptions of highly ionized material in GRB afterglow spectra might point to a special environment of material from the vicinity of the GRB itself (e.g., GRB 090426, \citealt{Thoene10}) and hence are not representative for the metallicity of GRB hosts in general. 

Extinction in GRB hosts is usually low, although somewhat higher than in QSO-DLAs. For a sample of 41 {\it Swift} GRB afterglow spectra, \cite{Zafar11} found that 90\% have $\textnormal{A}_V<0.65$ mag with an average of 0.24 mag and an SMC extinction law. A similar result was found from photometric data of {\it Swift} GRBs by \cite{Kann10}. However, high extinction makes the detection of the afterglow at optical wavelengths difficult and hence this result are somewhat biased (e.g., \citealt{Greinerdarkgrbs, KruehlerDusty}). The extinction in GRB hosts also seems to decrease  with redshift \citep{Kann10, Zafar11}. Again, we might miss more highly extinguished high-redshift bursts, due to the shape of the extinction curve which particularly affects optical observations, as these probe the rest-frame UV at those redshifts. Furthermore, dust is produced by AGB stars and SNe which, in a young star-forming region at high redshifts, might not be present or not have produced enough dust yet (see e.g. \citealt{Gall11rev}).

GRB 100219A was discovered in an image trigger by the {\it Swift} satellite \citep{Gehrels04} on 2010 Feb. 19, 15:15:46 UT \citep{RowlinsonGCN} and had a duration of T$_{90}=$ 18.8 $\pm$ 5.0 s \citep{BaumgartnerGCN}. An X-ray as well as an optical afterglow \citep{JakobssonGCN} was found. A nearby galaxy detected in the SDSS was first assumed to be the host galaxy \citep{BloomGCN}, implying a low redshift for this GRB. However, GROND (Gamma-Ray burst Optical/Near-Infrared Detector) reported a $g^\prime$-band dropout and suggested a redshift of $z\sim4.5$ \citep{KruehlerGCN}. We took spectra with the second-generation instrument X-shooter at the VLT, revealing a DLA system and a number of absorption lines at a common redshift of $z=4.6667$ \citep{GrootGCN,AntonioGCN}. This redshift was later confirmed by GMOS-N/Gemini-N \citep{CenkoGCN}.

In the following we present the observations and data reduction of the X-shooter spectra (Sect. 2), the analysis of the absorption lines in the host as well as the metallicities and abundances derived from fitting the column densities of the absorption lines (Sect. 3), a broad-band fit to the spectral energy distribution (SED) from X-ray to NIR (Sect. 4) and a description of the intervening system at $z=2.18$ (Sect. 5). Finally, we place the metallicity of GRB 100219A in the context of other GRB host metallicities and high-redshift galaxies in general (Sect. 6). For all calculations  we use a cosmology with $\Omega_\mathrm{M}=0.27$, $\Omega_\Lambda=0.73$ and H$_0=71$ km s$^{-1}$Mpc$^{-1}$. The flux of a GRB is defined to behave as $F_{\nu} \propto t^{-\alpha} \nu^{-\beta}$

\section{Observations}

\subsection{X-shooter spectroscopy}\label{specobs} 

We observed the afterglow of GRB 100219A starting on 2010 Feb. 20, 02:31 UT (12.5h after the trigger) with X-shooter at the VLT under good seeing conditions ($\sim$0.7 arcsec). The observations consist of 4 different exposures with an exposure time of 1200\,s each. The individual exposures were taken by nodding along the slit with an offset of $2^{\prime\prime}$ between exposures, using a standard ABBA sequence\footnote{The spectrum is taken at a pos. A in the slit, then shifted along the slit to pos. B etc.}. The spectra cover the wavelength range from $\sim3000$ to 25,000 {\AA} at a resolution of R$\sim\,$9,000--\,12,000 (visible/NIR range) and an average S/N of $\sim4$ per pixel. Due to the high redshift and the faintness of the source, we detect a continuum from the afterglow in the range from $\sim$ 5100 (Ly break) to 23,000 \AA. 

The spectra were processed with version 1.3.7 of the X-shooter data reduction pipeline \citep{Goldoni06, Modigliani11}.
The pipeline transforms the detected counts in physical quantities while propagating the errors and bad pixels consistently during the process. The reduction consists of the following steps. First the raw frames are subtracted from each other and cosmic ray hits detected and corrected using the method developed by \cite{vanDokkum01}. The frames were then divided by a master flat field produced using daytime
flat field exposures with halogen lamps. From the flat-fielded raw frames, the orders are extracted and rectified in wavelength space using a wavelength solution previously obtained from calibration frames. The resulting rectified orders are then shifted by the offset used in the observation and merged to obtain the final 2D spectrum, in overlapping regions, the orders are weighted by the errors.
 
From the resulting merged 2D spectrum we extracted a 1D spectrum together with a corresponding error spectrum and bad pixel map using a simple PSF weighting scheme under IDL. The error spectrum is derived from the noise of the raw frames propagated through all the reduction steps using the standard formulae of error propagation. However, probably due to the presence of the nearby galaxy, the absolute flux level of the 1D spectrum is  too low. Indeed in the UVB arm and in the strong atmospheric absorption bands in the NIR arm the median flux goes below zero by an amount slightly greater than the 1\,$\sigma$ error which is unphysical. We therefore applied the same positive offset to the flux in the three arms to correct for this effect and put the median flux to zero in those regions.

\begin{figure}
\centering
\includegraphics[width=\columnwidth]{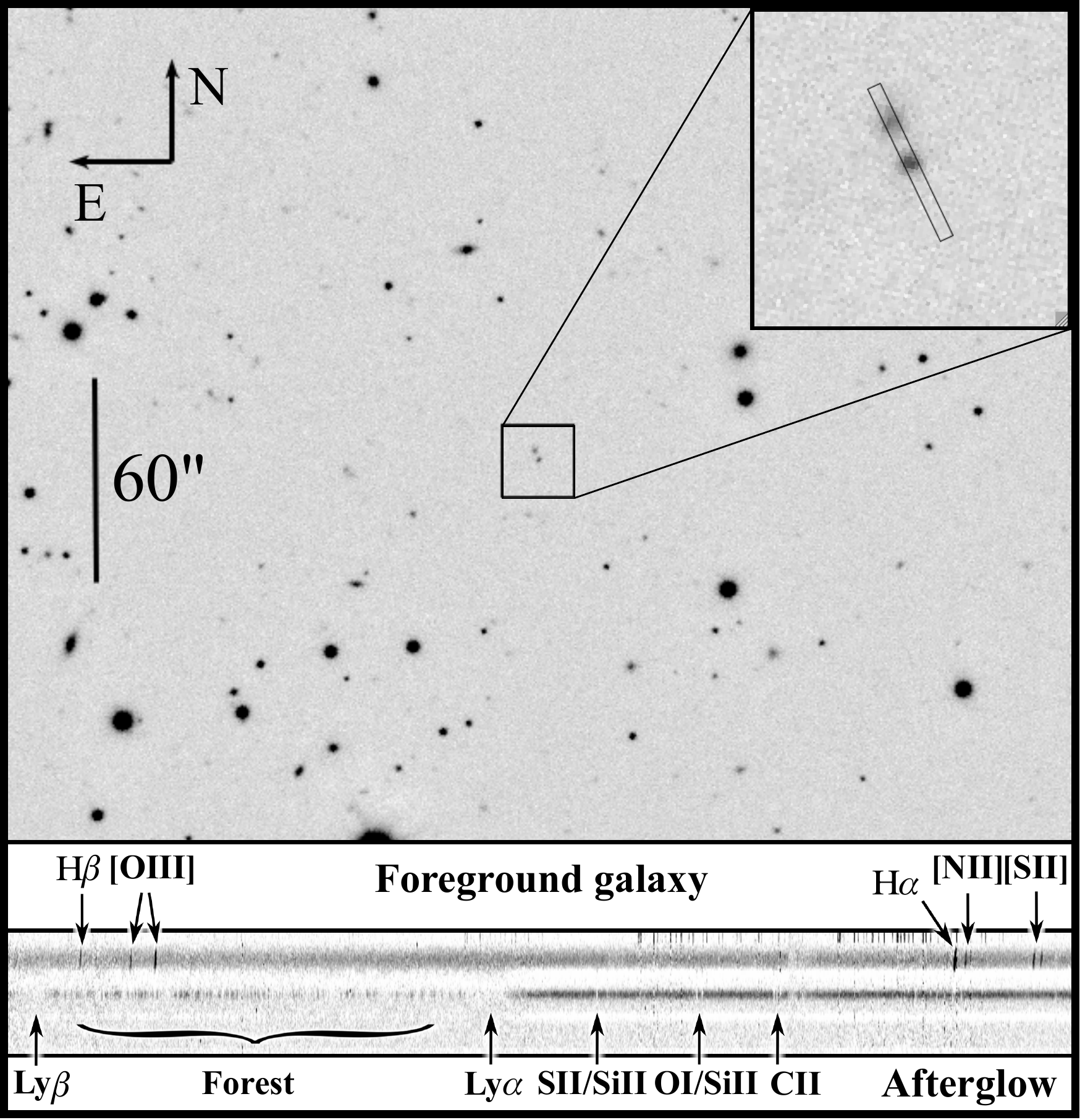}
\caption{Field of GRB 100219A in z-band from WHT/ACAM at 8.8\,h after the burst. The slit position across the afterglow and the nearby galaxy is indicated. The lower panel shows the 2D spectrum with the spectrum of the afterglow (lower trace) and the unrelated galaxy in the slit (upper trace) with some emission lines indicated. \label{fig:finder}}
\end{figure}

To perform flux calibration we extracted a spectrum from an observation of the flux standard GD71 \citep{Bohlin95} in staring mode. The reduction procedure is the same as for the science object, with the exception that the sky emission lines are subtracted according to the
Kelson (2003) method. The standard-star spectrum was then divided by the flux table of the same star obtained from the CALSPEC HST database\footnote{\cite{Bohlin2007}, http://www.stsci.edu/hst/observatory/cdbs/cal-\,spec.html} to produce the response function. After division by the flux table, we verified that the shape of the response function was compatible with the ones found in previous
reductions by our group. The response is then interpolated in the atmospheric absorption
bands in VIS and NIR and applied to the spectrum of the source. No telluric correction was performed. The final spectrum was transformed from air- to vacuum wavelengths. An image of the field including the slit position and a part of the 2D spectrum is shown in Fig.\ref{fig:finder}.

\subsection{Imaging observations}\label{sec:photometry}

The Gamma-Ray burst Optical/Near-Infrared Detector GROND \citep{GROND} mounted at the MPG/ESO 2.2~m telescope at La Silla observatory in Chile observed the field of GRB~100219A starting February 20, 2010, 00:30 UTC, 9.24~h after the BAT trigger \citep{KruehlerGCN}. Further imaging was performed on three later epochs with mean times of 36.6~h, 61.5~h and 110.7~h after $T_0$. In addition, we obtained two epochs of NIR observations in the $JHK_s$ filters with HAWK-I/VLT \citep{Hawki} at mean times of 59.6\,h and 107.8\,h  after the trigger, respectively. We also obtained a single epoch observation of the field using the Auxiliary
Port Camera (ACAM) on the 4.2 m William Herschel Telescope (WHT) on La Palma, starting 8.8\,h post-burst.  We used the $z$-band filter, with a total exposure of 1800\,s.  

\begin{figure}
\centering
\includegraphics[width=\columnwidth]{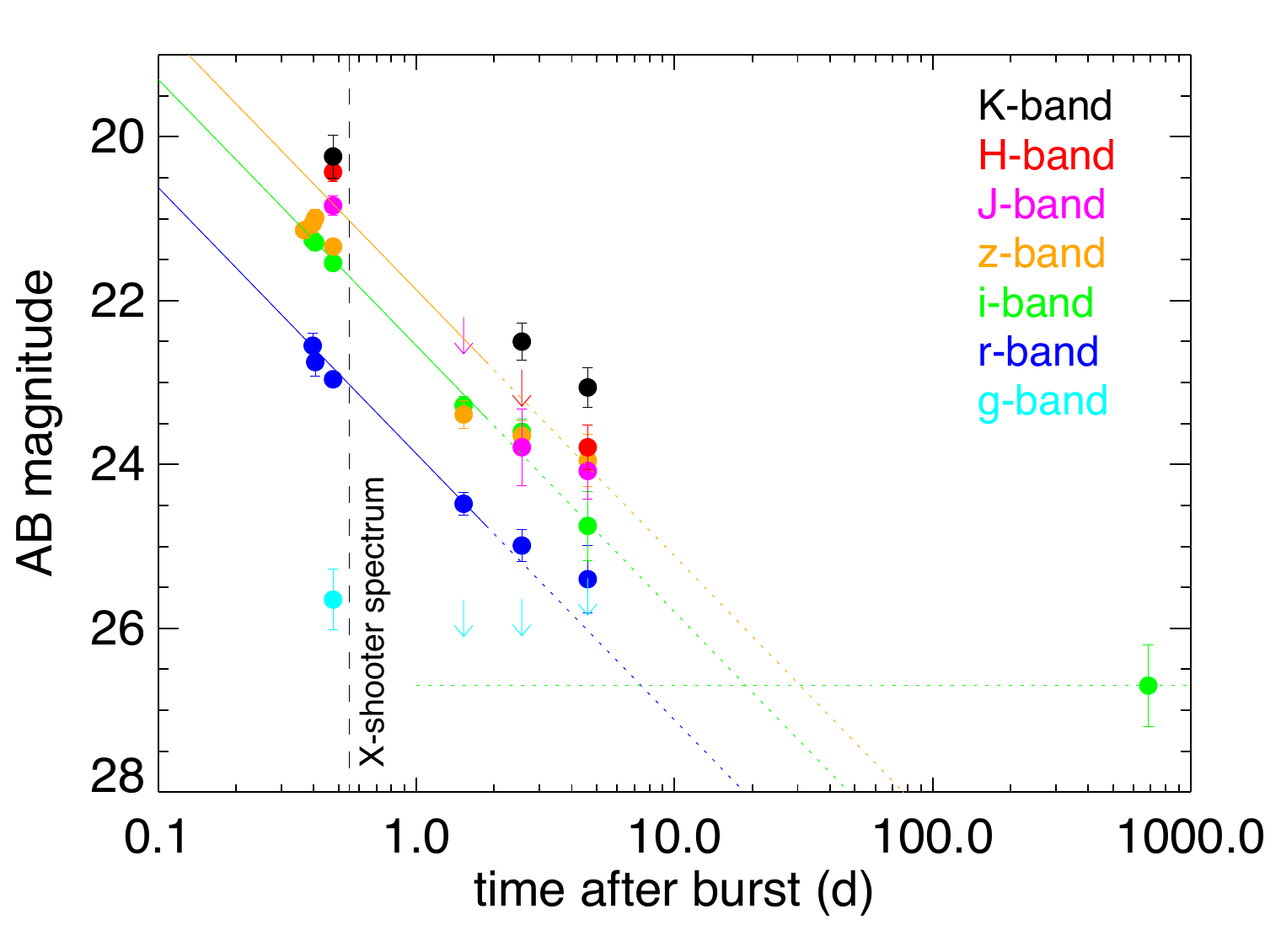}
\caption{Multicolor light curve of the late optical observations from GROND, HAWK-I and ACAM. Solid/dashed lines are the decay slope of $\alpha=-1.31$ as derived from a common fit to the $r^\prime$, $i^\prime$ and $z^\prime$ bands at the first four epochs. The following two epochs were excluded as there might be some rebrightening of the afterglow. The last point is the detection of the host galaxy. Magnitudes are in the AB system and corrected for Galactic extinction (see values in Table \ref{lightcurve}) \label{fig:lightcurve}.}
\end{figure}

All photometric data were reduced in a standard manner using pyraf/IRAF, similar to the procedure outlined in \citet{Kruehler08}. Given the presence of a bright foreground galaxy 3 arcsec from the afterglow, Point Spread Function (psf)-fitting photometry was used for the photometry in all epochs. The stacked images were flux calibrated against GROND observations of SDSS fields \citep{SDSS} taken immediately before or after the field of GRB~100219A for the optical $g' r'  i'  z'$, and magnitudes of 2MASS field stars \citep{2MASS} for the $JHK_s$ filters. All data were corrected for the Galactic foreground extinction of $E(B-V)=0.076$ according to \citet{Schlegel} assuming a total-to-selective extinction of $R_V = 3.08$. The WHT observations were calibrated via the GROND magnitudes for stars in the field. For a log of the observations and photometry see Table \ref{lightcurve}, the light curve is shown in Fig. \ref{fig:lightcurve}.

Fitting the afterglow light curve with all available optical/NIR data simultaneously with an achromatic power-law results in a temporal index of $\alpha = 1.31^{+0.03}_{-0.04}$ with a $\chi^2$ of 36 for 30 degrees of freedom (d.o.f.). There is a trend of flattening or rebrightening observed at late times ($t >$ 50~h) and we therefore exclude the last two epochs from the fit. \cite{Mao12} also found late achromatic bumps in the optical/X-ray lightcurve during the first day.

\begin{figure}
\centering
\includegraphics[width=\columnwidth]{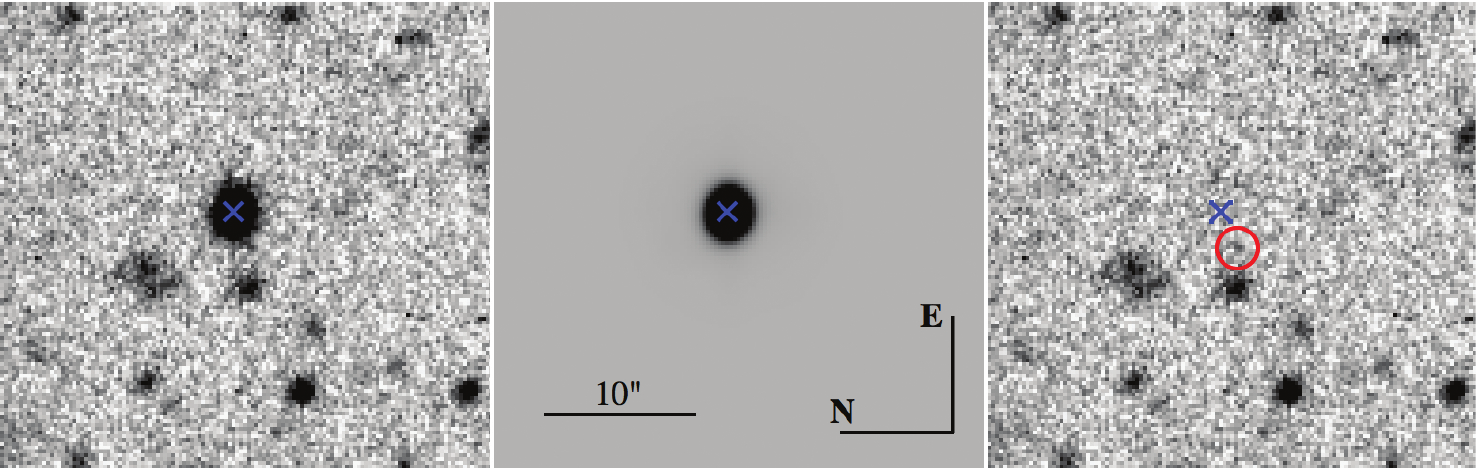}
\caption{Observations of the host galaxy candidate of GRB 100291A with OSIRIS/GTC, $i'$ band. Left panel: Combined $i'$-band observations. Middle panel: Model of the foreground galaxy, the center is indicated by a cross. Right panel: Image with the residuals after subtracting the galaxy model. A faint object is visible with $i'$=26.7$\pm$0.5 (circle). \label{fig:host}}
\end{figure}

To determine whether the late afterglow epochs might be dominated by a (bright) underlying host galaxy, we reobserved the field on Dec. 27 2011 in $i'$-band with OSIRIS/GTC and a total exposure time of 12$\times$120\,s. Since the GRB position lies near the bright foreground galaxy at $z=0.25$ also present in the X-shooter spectrum (see above), we use GALFIT \citep{Peng} to subtract the contribution of this foreground galaxy. The subtraction with a S\'ersic-profile works very well as shown in Fig. \ref{fig:host} and we detect a weak object at the position of the GRB with $i'=26.7\pm0.5$ (2 $\sigma$, AB magnitudes), using the same reference stars as for the afterglow photometry. The host galaxy candidate can therefore not have contributed significantly at 2--3 days post burst and the deviation from the powerlaw decay of the afterglow might be due to a late rebrightening.


\begin{table*}
\centering
\caption{Late time optical and NIR photometry. Times are midtimes of the observations compared to the burst trigger time. Observations in brackets are the stack of the three images above. Magnitudes are given in the AB system and corrected for Galactic foreground extinction of $E(B-V)=$0.076. Upper limits are $3\sigma$. \label{lightcurve}}
\begin{tabular}{lllllll}					
\hline\hline		
date	& exp. time & instrument& $g^\prime $ & $r^\prime $ & $i^\prime$ & $z^\prime$ \\
(h)&(s)&&(AB mag)&(AB mag)&(AB mag)&(AB mag)\\	\hline
8.812 	&1800&WHT/ACAM&---&---&---&21.14$\pm$0.017\\
9.51639 &	460  & GROND &---&	          $22.55\pm0.15$ & $21.26\pm0.09$ & $21.06\pm0.11$\\
9.73361 &	460  & GROND &---&	          $22.75\pm0.17$ & $21.29\pm0.07$ & $20.99\pm0.10$\\
10.9039 &	1460 & GROND &---&	          $22.88\pm0.05$ & $21.51\pm0.05$ & $21.32\pm0.06$\\
11.4039 &	1460 & GROND &---&	          $22.88\pm0.06$ & $21.55\pm0.04$ & $21.42\pm0.05$\\
11.9033 &	1460 & GROND &---&	          $22.99\pm0.06$ & $21.64\pm0.04$ & $21.44\pm0.07$\\
(11.4036&	4380 & GROND &$25.65\pm0.37$&	$22.96\pm0.04$ & $21.54\pm0.03$ & $21.34\pm0.04$)\\
36.5767 &	4380 & GROND &$>25.65$&       $24.48\pm0.14$ & $23.28\pm0.11$ & $23.39\pm0.17$\\
61.4692	& 4380 & GROND &$>25.64$&       $24.99\pm0.20$ & $23.60\pm0.15$ & $23.65\pm0.18$\\
110.692 &	4380 & GROND &$>25.39$&       $25.40\pm0.41$ & $24.75\pm0.42$ & $23.95\pm0.32$\\ \hline
685 days	& 1440  & OSIRIS/GTC&---&---&$26.7\pm0.5$&--- \\ \hline\hline
	&  &	& $J$ & $H$ & $K_s$ &\\ 
	&  & &(AB mag)&(AB mag)&(AB mag)&\\ \hline
10.9106 &	1200 & GROND &$20.78\pm0.17$&	$20.24\pm0.16$&---&\\
11.4106	& 1200 & GROND &$20.99\pm0.20$&	$20.32\pm0.16$&---&\\
11.9097	& 1200 & GROND &$20.75\pm0.16$&	$20.65\pm0.25$&---&\\
(11.4100& 3600 & GROND &$20.84\pm0.12$& $20.43\pm0.12$ & $20.24\pm0.26$)&\\
36.5831	& 3600 & GROND &$>22.22$	    &---&	---&\\
59.5956	& 2280 & HAWK-I&$23.79\pm0.47$	&$>22.84$	&$22.5\pm0.23$&\\
107.760	& 4560 & HAWK-I&$24.08\pm0.34$	&$23.79\pm0.27$	&$23.06\pm0.24$&\\
\hline	
 \hline
\end{tabular}
\end{table*}	

\section{Absorption lines in the GRB system}

We detect a number of resonant absorption lines from the GRB host galaxy, including a DLA, as well as one intervening system. The strongest absorption lines from the host system show at least 5 velocity components spanning a range of $\sim200$ km\,s$^{-1}$. We also detect two fine-structure transitions of \ion{Si}{II} and one from \ion{C}{II}, as have also been detected in a number of other GRB spectra (first detected by  \citealt{Vreeswijk04}). These transitions are likely produced by UV pumping of the excited levels by the radiation from the GRB itself. However, \ion{Si}{II*} has also been observed in star-forming galaxies where they could be produced by UV radiation from young stars (see e.g. \citet{Christensen10}). Finestructure lines that were so far only found in GRB environments such as \ion{Ni}{II*} or \ion{Fe}{II*} are not detected in our spectrum \cite{Christensen10, Vreeswijk07}.

\begin{figure*}
\centering
\includegraphics[width=17cm, angle=0]{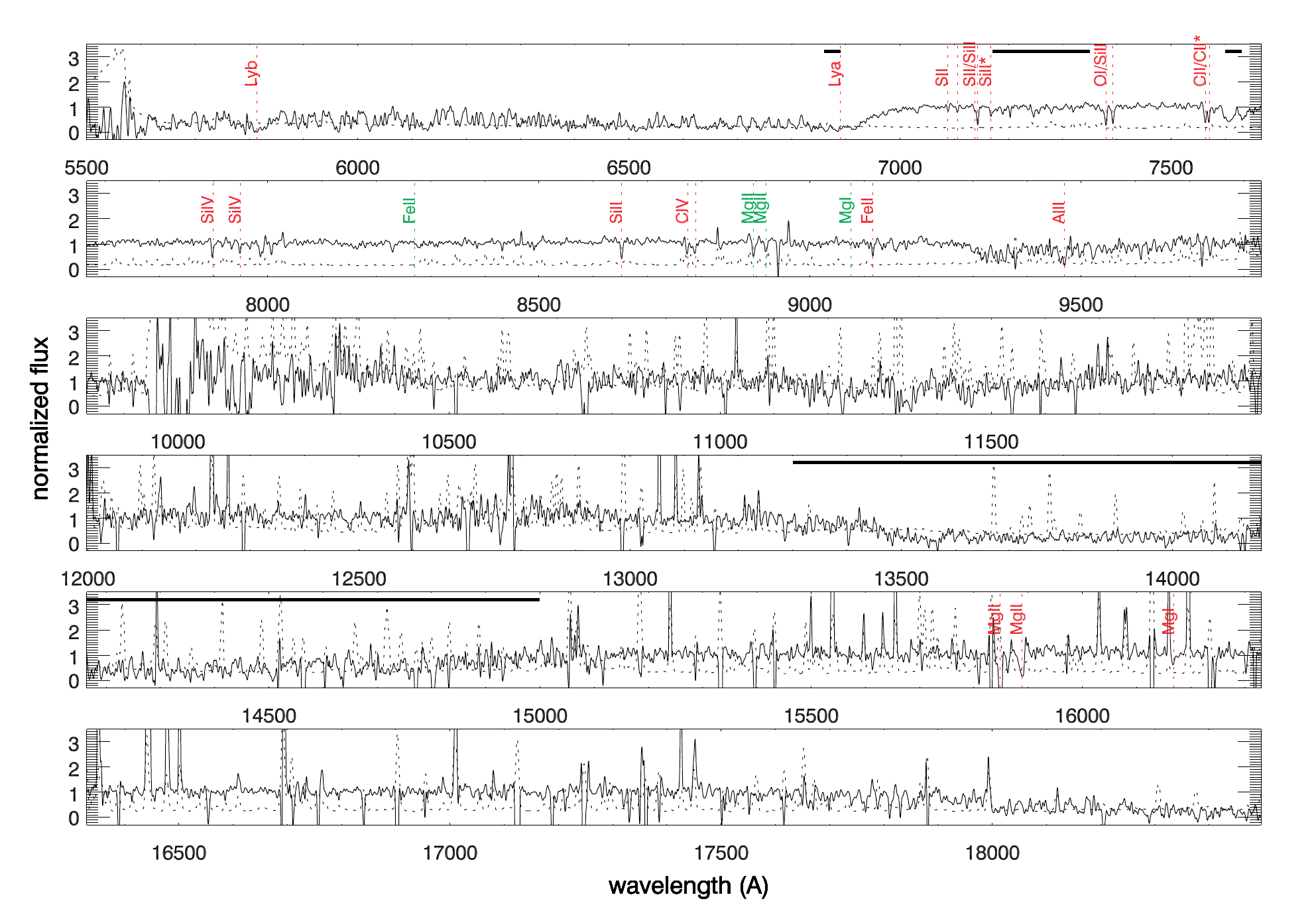}
\caption{X-shooter spectrum of 100219A with the detected absorption lines from the host system (red) and the intervening system (green) indicated. Horizontal black bars denote the atmospheric absorption bands, the dotted line shows the error spectrum. The UVB arm is not shown since it is below the Lyman limit. The spectrum above $\sim18,000$ {\AA} is not plotted due to low S/N and the absence of detected lines. \label{spectrum}}
\end{figure*}

\subsection{Line fitting and properties}\label{GRBlines}
From the absorption lines detected in the X-shooter spectrum \cite{AntonioGCN} had determined a redshift of $z=4.6667\pm0.0005$, which we refine in the following. We downgrade the spectrum to a resolution of $\sim2$ {\AA} to eliminate the velocity structure and fit  Gaussians to all well-detected absorption lines. We do not include the high-ionization lines such as \ion{Si}{IV} and \ion{C}{IV} as well as the fine-structure lines since those have a different velocity structure (see also Sect. \ref{sect:kinematics}). This gives us the redshift of the gas with the strongest absorption component, which coincides with the redshift of the main absorption component of \ion{S}{II}, the weakest absorption line detected. Wavelengths have been corrected for the heliocentric velocity shift of 5.26 km\,s$^{-1}$ at the time of our spectrum. The final redshift is $z=4.66723\pm0.00037$, which we adopt as $v=0$\,km\,s$^{-1}$.

The equivalent widths (EWs) are also determined from the Gaussians fitted to the lines in the smoothed spectrum. The errors on the EWs are based on the noise spectrum in the 95\% width interval of the fitted Gaussian. Errors from the Gaussian fit itself are not included. In Table \ref{lines} we list the detected lines with their corresponding wavelength and EWs, values are given in restframe.

\begin{table}[!hpt]
\caption{Absorption line list for the GRB and the intervening system. All wavelengths are in vacuum, EWs are given in restframe. The \ion{Mg}{II} lines of the GRB host system are severely affected by atmospheric lines, hence we cannot give exact values. \label{lines}}
\begin{tabular}{lllll}					
\hline\hline									
line	& {\bf $\lambda_\mathrm{rest}$ }& $\lambda_\mathrm{obs}$ & EW & log N\\
	& ({\AA})& ({\AA}) &  ({\AA}) & (cm$^{-2}$)\\ \hline
{\bf GRB}&&&&\\
Ly $\alpha$ & 1215.670 & 6894 & --- &$21.14\pm0.15$ \\
\ion{S}{II} & 1250.584 & 7090.375 &$0.12\pm0.04$  & $15.23\pm0.05$ \\
\ion{S}{II} & 1253.811& 7107.643&  $0.14\pm0.04$  & $15.23\pm0.05$\\
\ion{S}{II} & 1259.519 &7139.867& $0.10\pm0.04$ & $15.23\pm0.05$\\
\ion{Si}{II} & 1260.422 & 7145.059 & $0.52\pm0.04$ & 15.03$\pm$0.06 \\
\ion{Si}{II}* &1264.738 & 7170.786&$0.47\pm0.05$ & 13.57$\pm$0.06\\
\ion{O}{I} & 1302.169 &7381.448 & $0.67\pm0.04$& 16.82$\pm$0.10 \\
\ion{Si}{II} & 1304.370& 7394.606& $0.55\pm0.05$& 15.03$\pm$0.06 \\
\ion{C}{II} & 1334.532&7564.966 &$0.56\pm0.05$ & $15.12\pm$0.06\\
\ion{C}{II}* & 1335.663 &7571.421  &  $0.45\pm0.05$& 15.11$\pm$0.04\\
\ion{Si}{IV} & 1393.755 & 7900.031 &$0.40\pm0.05$ & 14.08$\pm$0.08\\
\ion{Si}{IV} & 1402.770 & 7950.811 &$0.34\pm0.05$ & 14.08$\pm$0.08\\
\ion{Si}{II} & 1526.708  & 8654.931  & $0.50\pm0.07$ & 15.03$\pm$0.06 \\
\ion{C}{IV} &1548.195 & 8775.541 &$0.38\pm0.08$  & 14.43$\pm$0.04\\
\ion{C}{IV} &1550.770 &  8789.557 & $0.41\pm0.04$ & 14.43$\pm$0.04\\
\ion{Fe}{II} & 1608.451 & 9118.284  & $0.33\pm0.04$& 14.70$\pm$0.03\\
\ion{Al}{II} & 1670.787  & 9471.844 & $0.79\pm0.07$& 13.84$\pm$0.05\\
\ion{Mg}{II}& 2796.352& (15852.8)  & --- & ($>13.72$) \\
\ion{Mg}{II}& 2803.530& 15893.554 &$\sim$1.4& $>13.72$\\
\ion{Mg}{I}& 2852.965&16167.345  & $0.45\pm0.08$ & $12.89\pm0.10$ \\ \hline
{\bf Interv.} &&&&\\
\ion{Fe}{II}&2600.173&8270.620&$0.10\pm0.04$& $13.13\pm0.12$\\
\ion{Mg}{II}& 2796.352&8895.826 &  $0.90\pm0.08$&  $13.69\pm0.10$\\
\ion{Mg}{II}& 2803.530&8918.049 &$0.50\pm0.10$& $13.69\pm0.10$\\
\ion{Mg}{I}& 2852.965& 9074.719&$0.10\pm0.05$& $12.07\pm0.14$\\
\hline	
 \hline
\end{tabular}
\end{table}

From the full-resolution spectrum we determine the column densities by fitting Voigt profiles to the different absorption components using the FITLYMAN context in MIDAS\footnote{http://www.eso.org/sci/software/esomidas/}. This program fits Voigt profiles with the atomic parameters for each transition by convolving them with the resolution of the instrument. Wavelengths of the individual components, column density and b-parameter (turbulent and thermal) can ideally all be obtained by fitting the absorption profile. The thermal b-parameter is set to a low value since temperature broadening is not expected to play a role for the observed cold/warm interstellar medium.

Due to the low S/N, limited resolution and blending of different components, we obtained the turbulent b-parameter of the resonant lines from fitting one wing of the different components in the element that, in each specific component, showed the least blending and/or saturation. The resulting b-parameters are: 11.2, 9.5, 10.2 and 6.5 km\,s$^{-1}$ for comp. I, II, III and V respectively and 6, 10, 10 and 5  km\,s$^{-1}$ for the high ionization lines. For comp. 0 and IV we take the average b-parameter of 12.6 km\,s$^{-1}$ (see A. de Ugarte Postigo et al. in prep.). 
b-parameters for Mg as well as the intervening system could be obtained directly from FITLYMAN. Note that the Mg absorption lines, as the only lines in the IR-arm of the spectrum, are better resolved than the lines in the VIS-arm due to the slightly higher resolution (see Sect. \ref{specobs}). The b-parameters from fitting blended, unresolved lines do not have a real physical meaning \citep{Jenkins86} and the low b-parameters obtained (after the spectrum had been convolved with the instrumental resolution) confirms that the absorption components are actually unresolved.

\begin{figure}[!hpt]
\includegraphics[width=8cm, angle=0]{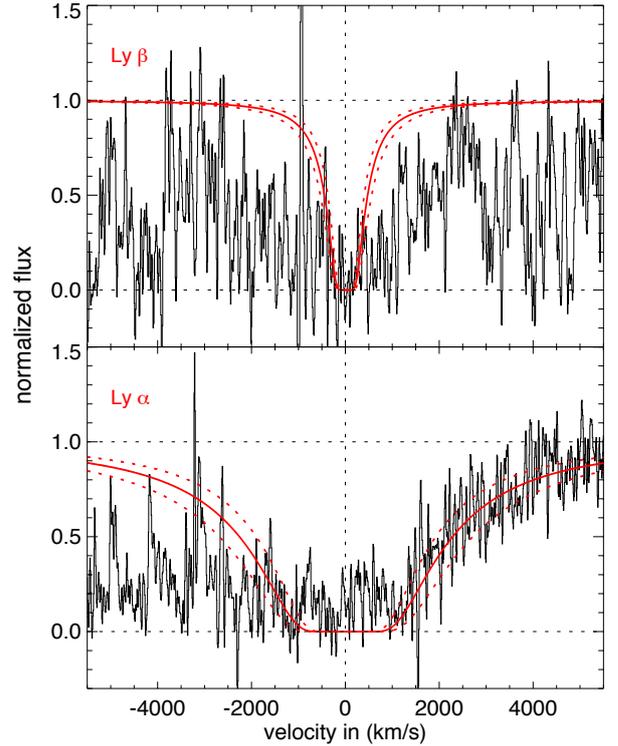}
\caption{Ly$\alpha$ and Ly$\beta$ absorption at the redshift of the GRB host. Ly$\beta$ was not fitted due to the Ly$\alpha$ forest, the fit shows the Ly$\beta$ Voigt profile with the parameters adopted from the fit to Ly$\alpha$.
\label{lyfit}}
\end{figure}

The column density of hydrogen was obtained from fitting the red wing of the Ly$\alpha$ line (see Fig.\ref{lyfit}). Ly$\beta$ is also in the range of the spectrum but cannot be fit since it is embedded in the Ly$\alpha$ forest and the Ly$\alpha$ forest blanketing is already playing a role at $z=$4.7. For \ion{Mg}{II}, we can only fit the $\lambda2803$\AA{} line of the doublet since the $\lambda2796$\AA{} line is heavily contaminated by atmospheric lines and even for \ion{Mg}{II} $\lambda2803$\AA{} we are only able to give fits for the three bluemost components. This is the highest redshift for which \ion{Mg}{II} has been detected in a GRB host and comparable to the highest strong (EW $>$ 1 {\AA}) \ion{Mg}{II} absorbers detected in QSO sightlines (e.g., \citealt{Jiang07}).

\begin{figure}[!hpt]
\includegraphics[width=8cm, angle=0]{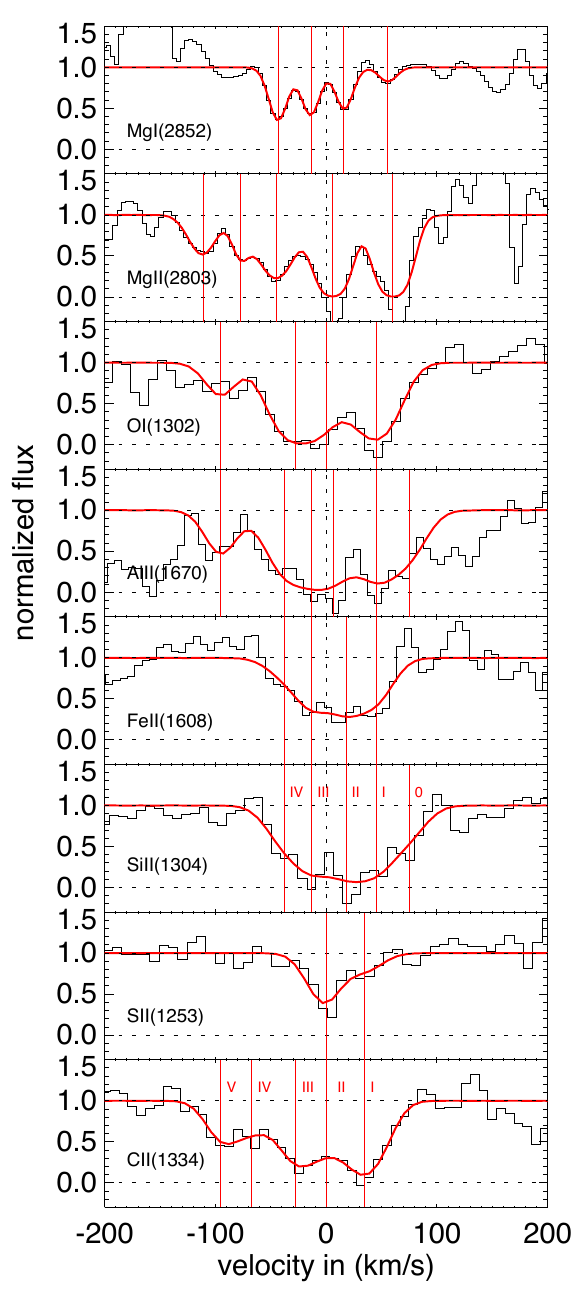}
\caption{Absorption components for the resonant transitions in the host of GRB 100219A. For Mg II, we show the fit to all components but do not consider the two redmost components for the column density due to contamination by the neighboring sky line.
\label{GRBlinefit}}
\end{figure}

\begin{figure}[!hpt]
\includegraphics[width=8cm, angle=0]{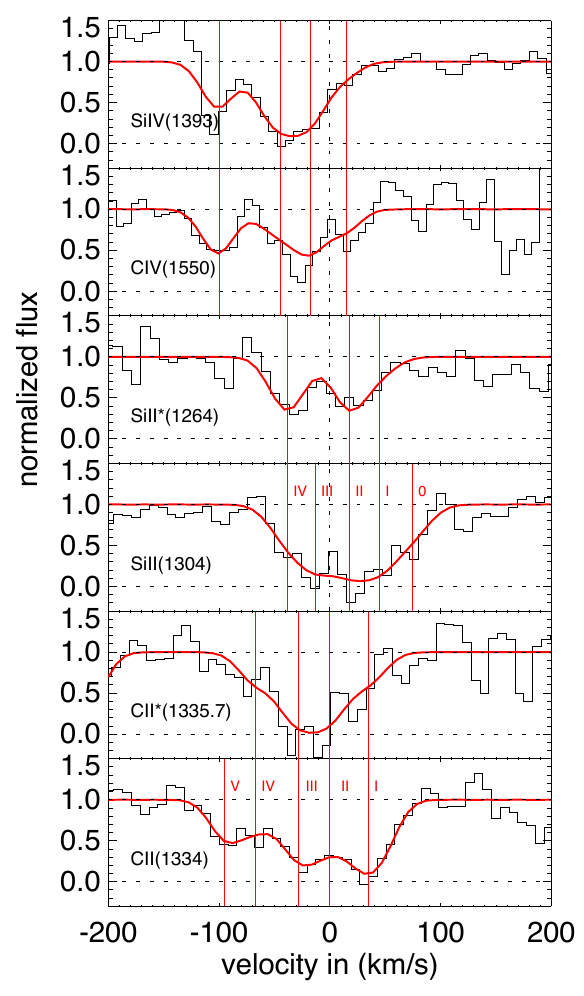}
\caption{Comparison between the absorption component of the low ionization, fine-structure and high ionization lines. While the low ionization lines have the strongest absorption in comp. I, it is comp. III -- IV for the high ionization lines and fine-structure transitions. Comp V is rather strong for the high ionization lines but is absent in the fine-structure and weaker in the resonant absorption lines.
\label{GRBlinefithigh}}
\end{figure}

In Table \ref{tab:components} we present the velocities and column densities for the different components. Fig. \ref{GRBlinefit} and \ref{GRBlinefithigh} show the Voigt profile fits to one transition from each ion, selected to either have the best S/N or being the least saturated transition or the one least affected by skylines and/or blending with other lines. The components are not at the same velocities for all lines which will be discussed in Sect. \ref{sect:kinematics}.\\

\begin{table*}[!hpt]
\caption{Column densities for the 5 velocity components in the absorption lines as shown in Fig.~\ref{GRBlinefit} and \ref{GRBlinefithigh}. Transitions in brackets are detected but affected by skylines or blended and have not been used to fit the column densities.}             
\label{tab:components}      
\centering                          
\scriptsize
\begin{tabular}{l | c c | c c | c c | c c | c c | c c }        
\hline\hline                 
 &    \textbf{ 0}  & \textbf{}  &    \textbf{ I}  & \textbf{} & \textbf{ II}  & \textbf{}  & \textbf{ III}  & \textbf{} & \textbf{ IV}& \textbf{} & \textbf{ V}\\    
  Ion,  Transition ({\AA})          & $\Delta$v &log N   & $\Delta$v &log N  & $\Delta$v&log N  &$\Delta$v&log N & $\Delta$v&log N & $\Delta$v&log N \\ 
      		&\scriptsize (km/s) &\scriptsize (cm$^{-2}$)&\scriptsize (km/s) &\scriptsize (cm$^{-2}$)&\scriptsize (km/s)&\scriptsize (cm$^{-2}$)&\scriptsize (km/s)&\scriptsize (cm$^{-2}$)&\scriptsize (km/s)&\scriptsize (cm$^{-2}$)&\scriptsize (km/s) &\scriptsize (cm$^{-2}$)\\
\hline\hline  
\ion{S}{II} 1250/53/59	&  ---	& --- & +35  & $14.42\pm0.24$	&   0	& $15.16\pm0.18$ &---	& ---	& --- & ---\\
\ion{C}{II}  1334 &  ---	& --- & +35  &   $14.99\pm0.37$  & 0  &   $13.78\pm0.26$ &  --28  & $14.24\pm0.28$ & --67  & $13.60\pm0.21$ & --95  &  $13.87\pm0.36$  \\ %
\ion{C}{II*}   1335.6/35.7 &  ---	& --- & +35 & $13.62\pm0.21$ & 0 & $14.81\pm0.53$ & --28 & $14.76\pm0.27$ & --67 & $13.62\pm0.18$ & --- & --- \\
\ion{C}{IV}  1548, 1550 &  ---	& --- & --- & --- & +15 & $13.49\pm0.19$ & --18  & $13.92\pm0.17$& --45 & $13.58\pm0.13$ & --100 & $14.07\pm0.19$\\
\ion{Si}{II}     1304 (1260, 1526)&  +75	& $13.79\pm0.33$ & +45  &  $14.44\pm0.48$ & +18  &$14.59\pm0.45$ &--13  & $14.41\pm0.35$&--38 &$13.91\pm0.18$&---&---\\
\ion{Si}{II*} 1264 &  ---	& --- &  +45 & $12.38\pm0.32$  &   +18 & $13.22\pm0.26$   &---  & ---& --38 & $13.25\pm0.19$ & --- & --- \\
\ion{Si}{IV}  1393, 1402  &  ---	& --- & --- & --- & +15 & $12.53\pm0.42$ & --18  & $13.62\pm0.34$& --45 & $13.68\pm0.18$ & --100 & $13.44\pm0.23$\\
\ion{Fe}{II}   1608, (2344) &  --- & --- &+45  & $14.11\pm0.15$ & +18& $14.25\pm0.22$& --13  &  $14.20\pm0.20$  & --38 & $13.55\pm0.27$ & --- & ---\\
\ion{O}{I}   1302 &  ---	& --- & +45  &  $15.93\pm0.43$ & 0  & $14.67\pm0.43$& --28 & $16.76\pm0.56$  & --- & --- &  --95 &  $14.18\pm0.25$\\
\ion{Al}{II} 1670 &  +75	& $12.65\pm0.19$ & +45  & $13.21\pm0.39$& +7  & $13.32\pm0.48$ & --13 & $13.00\pm0.65$& --38 & $12.96\pm0.26$ & --95 & $12.88\pm0.36$ \\
\ion{Mg}{I}  2852&  ---	& ---  & +55 &$11.47\pm0.49$ & +16& $12.24\pm0.36$  & --13 & $12.35\pm0.34$  & --43 &$12.54\pm0.38$ & --- & --- \\
\ion{Mg}{II}  2803, (2796) &  (+60 & $14.99\pm0.79$) & --- & ---  & (+6 & $14.74\pm0.70$) & --45 & $13.38\pm0.14$ & --77 &$13.28\pm0.63$  & --111 & $12.94\pm0.17$\\
\hline\hline
\end{tabular}
\end{table*} 
 
\subsection{Metallicity}\label{metallicities}

The total column densities of different absorption lines allows us to determine the metallicity of the galaxy hosting GRB 100219A. Leaving out Al and Mg which are affected by atmospheric lines we get the following metallicities:  $\textnormal{[C/H]}=-2.1\pm0.1$, $\textnormal{[O/H]}=-1.0\pm0.1$, $\textnormal{[Si/H]}=-1.5\pm0.1$, $\textnormal{[S/H]}=-1.0\pm0.1$ and $\textnormal{[Fe/H]}=-1.9\pm0.1$. For C and Si we added the contribution of the low- and high-ionization resonant and fine-structure transitions. 

\begin{figure}[!hpt]
\includegraphics[width=\columnwidth]{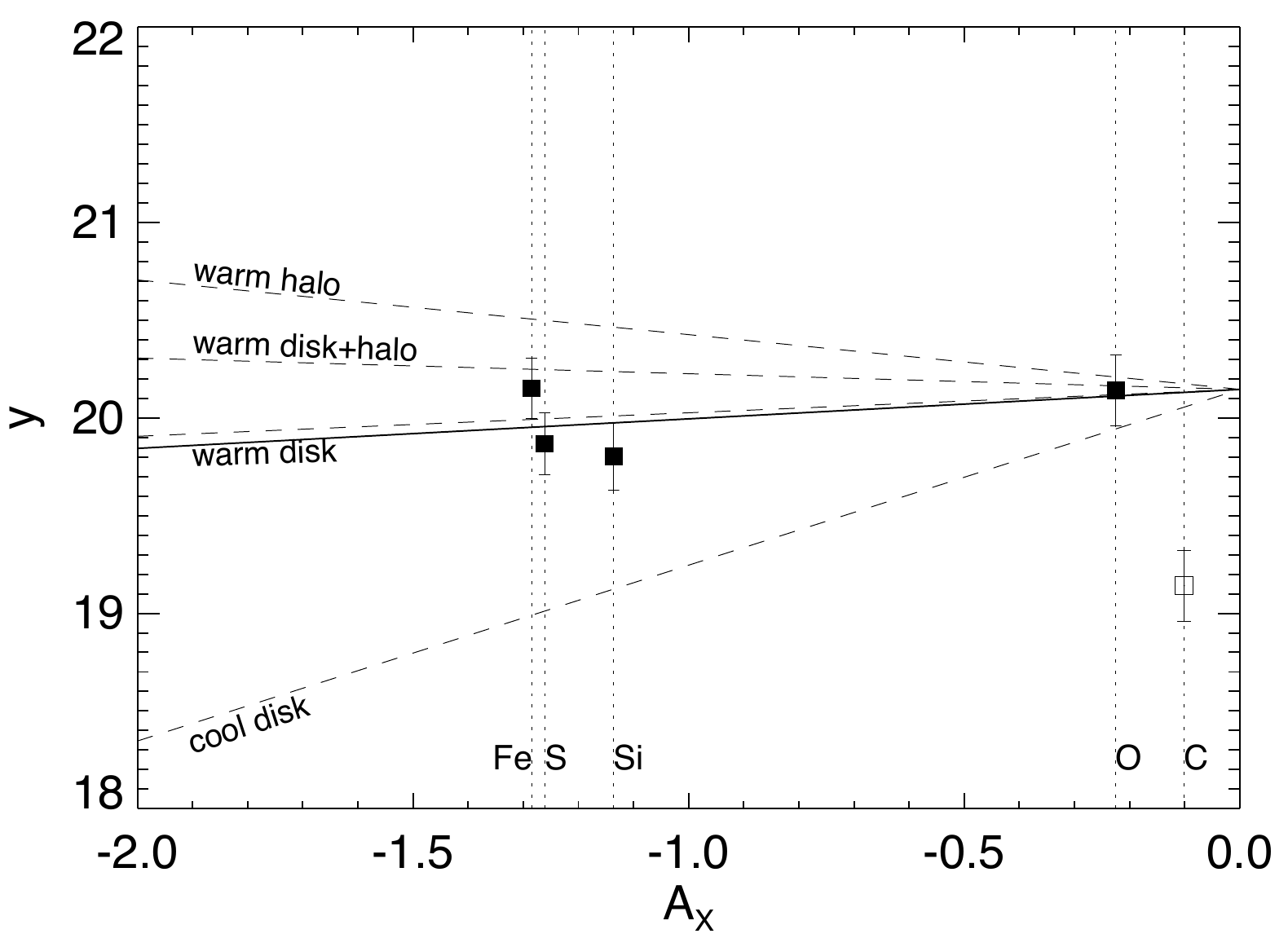}
\caption{Abundances measured in GRB 100219A and plotted according to the dust depletion method described in \cite{Jenkins09}. A$_x$ gives the propensity of an element to convert from gas to solid which roughly correlates with element condensation temperature and it hence fixed for each element. y shows the column density convolved with several parameters that \cite{Jenkins09} obtained by fitting several sightlines in the MW such that for a certain depletion 
 along a sightline, the elements lie on a straight line with slope F* (the ``depletion factor''). The metallicity is then obtained by the difference between log N$_\mathrm{H}$ and the intersection of the linear fit with the y-axis (the normalization). The solid line shows the linear fit to our data, resulting in a depletion factor of F*$=$0.16 and a metallicity of $\sim$--1.0. The dashed lines show abundance pattern according to different types of sightlines as described in \cite{Savage96}, assuming the same metallicity of --1.0. C was not included in the fit. \label{fig:depletion} }
\end{figure}

O and S are the elements known to be least affected by dust depletion but OI is usually saturated and the column density might not be fully reliable. However, for our spectrum they both show similar metallicities. \cite{Jenkins09} investigated the depletion for a range of different elements in sightlines through the MW. With the fits obtained for those elements, they developed a method to determine the degree of depletion $F*$ and the metallicity by a linear fit of the individual behaviour of the different elements in case of depletion. The normalization of the linear fit then gives a synthetic hydrogen abundance, whose difference to the measured \ion{H}{I} column density (if available) gives the metallicity. Applying this method for our spectra results in a mild depletion ($F*= 0.15$), leaving out the low abundance from C, a depletion pattern similar to a warm disk+halo in the MW (see also \citealt{Savage96}) and a metallicity of around -1.0. In Fig.\ref{fig:depletion} we plot the values from GRB 100219A according to the method of \cite{Jenkins09} together with the depletion factors of different patterns from our MW as defined in \cite{Savage96}. 
 
\begin{figure}[!hpt]
\includegraphics[width=\columnwidth]{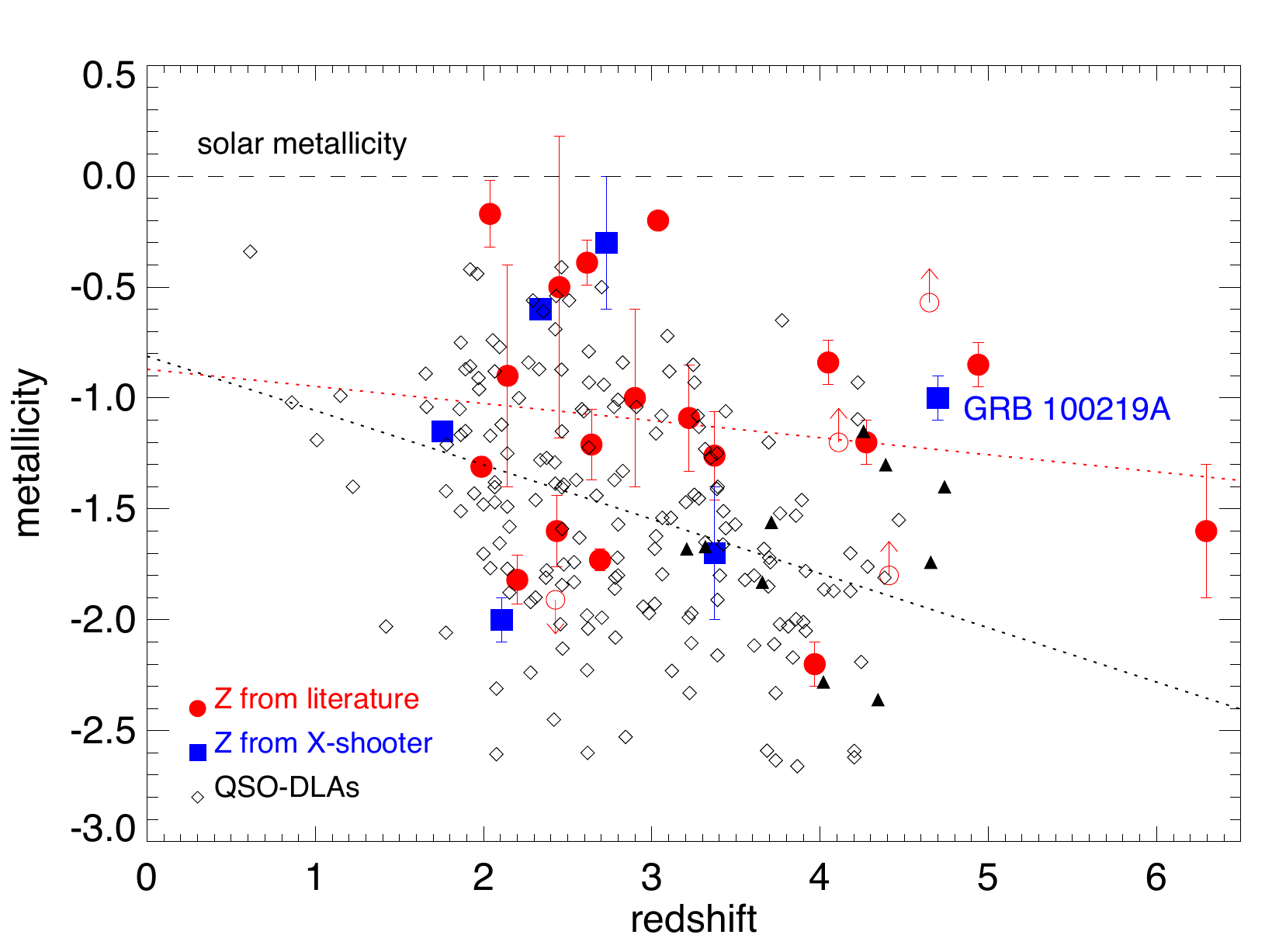}
\caption{Currently published GRB host metallicities from absorption lines (filled dots) including X-shooter metallicities (filled squares) compared to QSO-DLA metallicities (diamonds and filled triangles), error bars are omitted for clarity. The GRB values are taken from \cite{Fynbo06} including later published measurements: GRB 060223A and GRB 060510B \citep{Chary}, GRB 060526 \citep{Thoene10}, GRB 070802 \citep{Eliasdottir09}, GRB 071031, GRB 080310 and GRB 080413A \citep{Ledoux09}, GRB 080210 \citep{DeCia11}, GRB 080607 \citep{ProchaskaH2}, GRB 090205 \citep{DAvanzo10}, and GRB 090516 (A. de Ugarte Postigo in prep.). For GRB 050904 ($z=6.29$) we adopt a new, lower value, as described in Sect. \ref{sect:z4grbs}. The values from X-shooter are from \cite{DElia10} (GRB 090926A), M. Sparre et al. in prep. (GRB 090809, GRB 100425A) and \cite{Antonio090313}/J. Fynbo (priv. comm.) for GRB 090313. QSO-DLA metallicities are from \cite{ProchaskaDLA} (diamonds)  and \cite{Rafelski12} (filled triangles) \label{fig:metal}. }
\end{figure}

Throughout the paper, we will adopt the metallicity of S since it is known to be least affected by dust depletion and therefore assume a metallicity  $\textnormal{[M/H]}=-1.0\pm0.1$ or $\sim1/10$ Z$_\odot$ for GRB\,100219A. This metallicity lies well within the value found for other GRB hosts at lower redshifts (see Fig. \ref{fig:metal}). In fact, GRB host metallicities in the range $3<z<4$ show a large scatter between $<1/100$ Z$_\odot$ (GRB 050730, \citealt{DElia07}) and $>0.3$ Z$_\odot$ (GRB 090205, \citealt{DAvanzo10}, which has only a slightly lower redshift than GRB 100219A). QSO-DLA metallicities at this redshift range are considerably lower with a mean metallicity of around 1/100 Z$_\odot$ . Measurements from QSO absorption systems at $z=5-6$ \citep{Becker11} indicate an even lower metallicity for those galaxies. We will discuss this further in Sect. 6.2.

\subsection{Kinematics}\label{sect:kinematics}
The strongest absorption lines cover a velocity range of around 150\,km\,s$^{-1}$, but with different components and distributions. \ion{C}{II}, \ion{O}{I} and \ion{Al}{II} show the largest velocity spread while \ion{Si}{II} and \ion{Fe}{II} are narrower and have a somewhat different distribution. \ion{Mg}{II} has a completely different velocity distribution and might trace different material than the other resonant lines.

While for GRBs high and low ionization lines often show the same velocity range, the opposite is true for QSO absorbers where high and low ionization lines trace different volumes (ionized halo gas vs. low ionization gas in the central parts of the galaxy). The situation is different for GRB 100219A.
The high ionization lines of \ion{Si}{IV} and \ion{C}{IV} are blueshifted compared to their low ionization counterparts but have a similar distribution of components. The mean absorption of the high ionization lines is at $v=-50$\,km\,s$^{-1}$ while it is $v=0$\,km\,s$^{-1}$ (by definition) for the low ionization lines. This also implies a different ionization of the material in the individual velocity components with a very low ionization for comp I and II, the strongest components of the low ionization lines, provided the absorption actually arises in the same place for low and high ionization lines. 

The situation is equally puzzling comparing the fine-structure lines. The fine-structure lines of \ion{Si}{II*} and \ion{C}{II*} show a similar velocity width than the resonant counterparts but the strongest absorption component of the fine-structure line does not coincide with the main absorption (or the redmost) component and lies at around $-40$ km\,s$^{-1}$. Provided that the fine structure lines originate from the UV radiation of the GRB, the strongest fine-structure absorption should be closest to the GRB site and is therefore often found to be in the redmost component (see, e.g., \citealt{Thoene08}), as expected from an ordered velocity field in the host galaxy. If the GRB radiation is indeed responsible for the excitation of our fine-structure lines, component III must be the one closest to the GRB as we will show in the following.

Using the ratio between fine-structure and resonant abundances of the same element, one can determine a rough distance of the material from the GRB itself, assuming the levels are populated by indirect UV pumping from the GRB (see \citealt{Prochaska06finestruc, Vreeswijk07}). We compute the distance between the GRB and the absorbing systems using a time-dependent, photo-excitation code (see \citealt{DElia07, Vreeswijk07} for details) predicting the ratio between fine structure and ground state levels. We use as input the GRB flux, spectral and temporal indices estimated in this paper, a Doppler parameter of $12.6$ km/s (see section 3.1), and we assume that the gas is completely in its ground states when the GRB explodes. The output of our code is then compared to the \ion{Si}{II} column densities for the components that also show absorption from \ion{Si}{II*}, i.e., components I, II and III. The initial SiII ground state column densities is obtained by summing the contribution from \ion{Si}{II} and \ion{Si}{II*} for each component. Given the large error bars of our data, we only derive an order of magnitude for the GRB-absorber distances. We find that component IV is the closest to the GRB, at a distance of d$_\mathrm{IV}\sim 300$ pc and a similar distance for comp. II while component I is at d$_\mathrm{I} \sim 1$ kpc.

The complicated velocity structure of the different absorption species could be an indication for a chaotic velocity field in
this galaxy. High-redshift star-forming galaxies are likely nothing close to the ordered, rotation-supported galaxies
we see today. IFU observations of galaxies at $z\sim2-3$  show a high velocity dispersion of the measured emission lines \citep[see, e.g.][and references therein]{Law09} whose origin is still highly debated. Among other mechanisms, mergers \citep{Puech06, Puech07} and massive star formation \citep{Lehnert09} have been proposed. \cite{FoersterSchreiber} conclude from galaxies
in the SINS survey that around 30\% of the galaxies at z$=$1--3 are rotationally supported, but turbulent disks probably
indicate they have completed most of their formation process, while another 30\% might be galaxy mergers. However,  \cite{Law09} suggest that clumping of gas during the formation process and subsequent infall to the center of the galaxy can explain
the observed velocity field and mimic the appearance of a velocity field produced by a merger. On the other hand the reverse is true
as \cite{Robertson06} show that a gas-rich merger can mimic a velocity field resembling to a disk rotation. Both merging and gas infall has to play an even larger role for galaxies at $z\sim5$.

The complicated velocity field observed for the host of GRB 100219A might either be explained with the infall of metal rich
material onto the galaxy or with the galaxy being involved in a large merger with another
galaxy. Infalling gas as proposed for the velocity structure of other high redshift galaxies should, according to the models, be largely pristine gas, hence this scenario seems less likely for our galaxy and the velocity field might indeed point to an ongoing merger. The large velocities observed for \ion{Mg}{II} could also indicate some sort of galactic outflow, like it has been found for
GRB hosts at low redshifts \citep{Thoene07}. Future studies have to show whether the velocity structure of the ISM in high redshift
galaxies and GRB hosts is more complex than the one of lower redshift galaxies.

\section{SED from X-ray to IR}\label{sect:SED}
The broad wavelength coverage of X-shooter also enables the derivation of the absolute extinction curve in combination with the X-ray data assuming that the intrinsic afterglow spectrum is a powerlaw or a broken power-law \citep[see e.g.][]{Watson12}.
The flux calibration of our spectrum can be considered reliable between 7000 and 23,000 {\AA}, beyond that, the flux of the afterglow is very low and we have to disregard this part of the spectrum. Below 7000 {\AA}, the spectral flux is decreased due to the Ly $\alpha$ forest. As a consistency check for the flux calibration of the spectrum, we also compare the continuum with the fluxes obtained from our GROND photometry at the time of the spectrum (midtime around 13.2\,h after the burst). The spectral continuum has been raised by a factor of 1.5 to match the photometric data which can be explained by slit losses. As found by our group in previous reduction (see e.g. Sparre et
al. 2011), the relative flux calibration is reliable even at these
flux levels while the absolute calibration may need to be adjusted
with photometry.

In order to cover a broader wavelength range, we also include
the X-ray spectrum from XRT on {\it Swift}. The XRT began
observing the field at 15:18:45.0 UT, 178.5 seconds after
the BAT trigger \citep{RowlinsonGCN}. The {\it Swift} XRT
observed GRB 100219A from 182 s to 88.5 ks for a total of
8.3 ks. The data are mainly in Photon Counting (PC) mode, plus 23 s in
Windowed Timing mode.
The light curve can be modeled with an initial power-law decay
with an index of $\alpha_ 1=1.53^{+0.28}_{-0.22}$, followed by a
break at 785 s to $\alpha_2=0.57\pm0.06$
and a final steepening at $\sim 4\times10^4$ s \cite{RowlinsonGCN2}.
A flare is present around 1300 s and another smaller one at $\sim 2\times 10^4$ s.

We extract a spectrum using the {\it Swift} XRT spectrum repository
\citep{Evans} in the time interval 5000--80000 s, during
which the X-ray spectral hardness ratio remained constant
(and therefore we do not expect strong spectral variations), resulting in 5.7 ks effective
exposure time. The resulting X--ray spectrum is rebinned to have at least 7 counts
per energy bin and has been fit using the Churazov weighting scheme within
the spectral fitting package XSPEC (v.12.7.0). The latest XRT
response matrix was used (v.013). We model the observed spectrum
with an absorbed power law model. Our fit includes two absorbers, one Galactic  (fixed
to $6.5\times10^{20}$ cm$^{-2}$,  \citealt{Kalberla}) and one intrinsic to the
GRB host galaxy at $z=4.667$ (we adopt the {\tt phabs} model).
The resulting fit is good with a reduced $\chi^ 2_{\rm red}=
1.1$ for 33 degrees of freedom. The photon index is $\Gamma=1.57^{+0.2}_{-0.13}$ ($90\%$ confidence level for one
interesting parameter), the intrinsic column density is $N_{HX}(z) = (3.1^{+3.9}_{-2.7})\times10^{22}$
cm$^{-2}$. The mean 0.3--10 keV (unabsorbed) X--ray flux during the select interval is
$ 2.5 \times 10^{-12}$ erg cm$^{-2}$ s$^{-1}$.

\begin{figure}[!hpt]
\includegraphics[width=\columnwidth, angle=0]{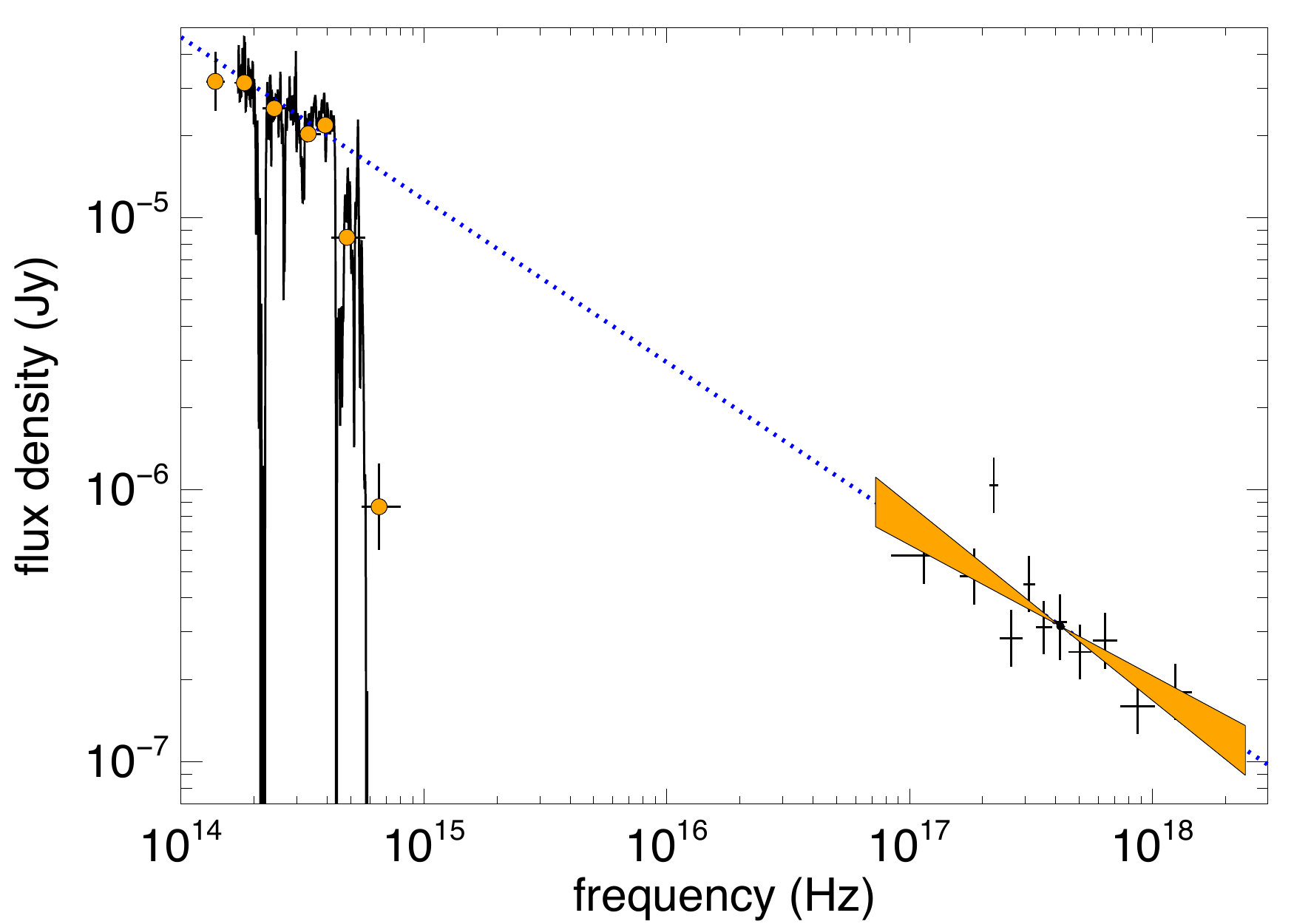}
\caption{Broad-band SED fitting of the X-shooter optical-NIR and X-ray spectrum. The black line shows the X-shooter spectrum corrected for an extinction of A$_V=0.24\pm0.06$ mag using an SMC extinction law. Orange filled symbols overplotted on the optical/IR spectrum are photometric points from GROND (see Table \ref{lightcurve}). The spectral slope with errors from the X-ray data is indicated by the shaded region. The extrapolation of the optical slope is consistent with the X-ray spectrum (dotted line). X-rays and optical data have been scaled to the corresponding flux at the time of the observations of the X-shooter spectrum using interpolations of photometric data from the X-ray and optical light curves.
\label{fig:ext}}
\end{figure}

The best fit to the X-shooter continuum results in an extinction 
of $A_V=0.24\pm0.06$ mag and an SMC extinction law  (see Fig.\ref{fig:ext}). The moderate extinction from the afterglow continuum is in line with the moderate dust depletion derived in Sect. 3.3. Dust depletion seems to be common in GRB hosts (see, e.g., \citealt{SavaglioNJP}) and higher than in QSO-DLAs despite the generally low extinction found from SED fits to the afterglow, the reason for which is still debated. The slope of the spectral energy distribution is $\beta_{opt}=0.60\pm0.12$ consistent
with the slope from X--rays of $\beta_{X}=0.57$ ($\beta=\Gamma-1$) \citep[see also][]{Mao12}.
To obtain a reasonable value for the electron distribution index $p$, this slope implies that the cooling break $\nu_c$, according to the fireball model (see e.g., \citealt{Sari}) has to lie bluewards of the X-shooter spectrum and $\beta=(p-1)/2$ with p$=$2.2. Fixing the
slope in the X--rays to the one from the optical spectrum, we get
an intrinsic absorption of $N_{HX}(z) = (5.1^{+3.7}_{-2.3})\times10^{22}$
cm$^{-2}$ (90\% confidence level) and $\chi^ 2_{\rm red}=1.1$. 

The column density derived from X-rays is about one order of magnitude higher than the column density of HI as derived from the optical of log N$=$21.14 (see Table \ref{lines}). In fact, the X-ray column density is above the average for {\it Swift} GRBs but comparable to the mean for $z>4$ GRBs of log N$_\mathrm{HX}\sim22.5$ following the trend of increasing column density with redshift \citep{Campana10}. This is partly an observational effect (at higher redshifts, a larger column density is needed for it to even be detected) but probably also due to an increased in intervening absorbers contributing to the total absorbing column. Gas-to-dust and metal-to-dust ratios for GRB 100219A are high as it is generally the case for GRBs, \citep{Watson12}, above the values for the local group \citealt{Zafar11}. This is usually explained by ionization of HI and in particular dust destruction by the GRB \citep{Watson07, Campana10, Fynbo09}. This could also explain the very large difference between optical and X-ray column densities (about a factor of 10), as generally found for GRBs, since X-ray absorption is not affected by ionization and hence most metals are found in highly ionized states due to the absence of a large column density of moderately ionized metals \citep{Schady11}.

\section{The intervening system}

Despite the high redshift of the GRB, we only detect one intervening system at a redshift of $z=2.180$ which might be a consequence of the low S/N of the spectrum. A second intervening system at $z=1.8$ that had been claimed in \cite{AntonioGCN} could not be confirmed in our refined analysis. We are not able to detect any possible absorption lines from the nearby galaxy at $z=0.25$. The redmost absorption transitions from the ISM commonly observed are Ca II $\lambda\lambda$ 3933, 3969 {\AA}, which, at $z=0.25$ still lie within the Ly$\alpha$ forest of the GRB absorption system, the weaker NaD doublet $\lambda\lambda$ 5890,5895 \AA is in a part of the spectrum affected by skylines.

\begin{figure}[!hpt]
\includegraphics[width=8cm, angle=0]{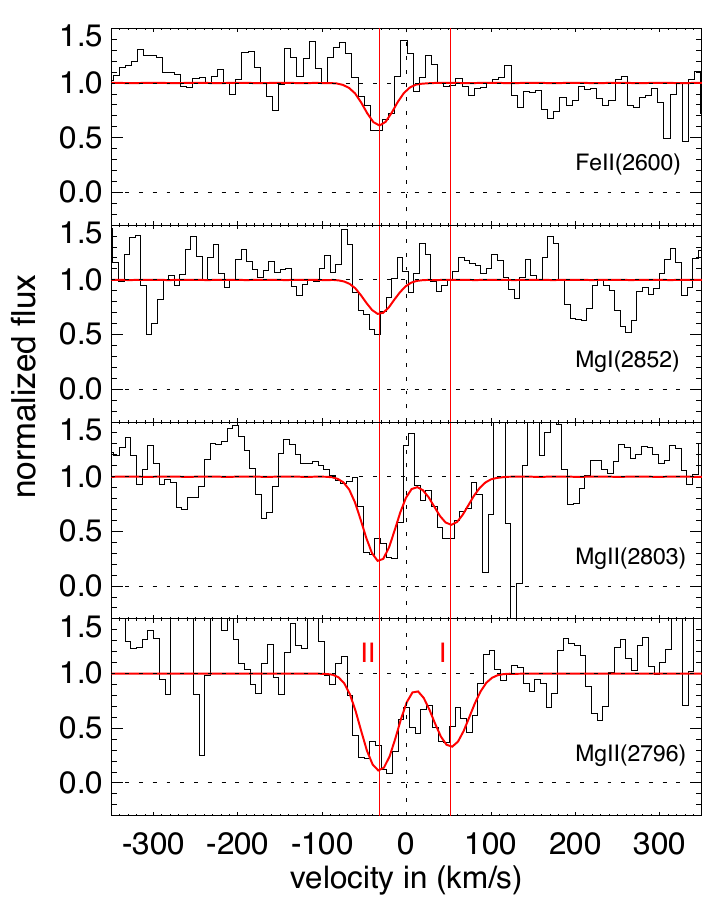}
\caption{Fits to the two velocity components in the intervening system at $z=2.181$. 
\label{GRBintervfit}}
\end{figure}

The refined redshift for the intervening system according to the method described in Sect. \ref{GRBlines} is $z_\mathrm{interv}=2.18116\pm0.00034$. For the $z=2.181$ system, we securely detect the \ion{Mg}{II} doublet $\lambda\lambda$ 2796, 2803, \ion{Mg}{I} and \ion{Fe}{II} $\lambda2586$, whereas the red part of the \ion{Fe}{II} $\lambda2600$ transition is affected by a skyline. \ion{Mg}{II} shows two components separated by 80 km\,s$^{-1}$ while \ion{Mg}{I} and \ion{Fe}{II} only show the stronger of the two components, likely due to the lower line strength of those transitions (see Fig. \ref{GRBintervfit} and Table \ref{tab:interv}).

The redshift path covered by our spectrum is of $\Delta z\sim2.6$. In the past (e.g., \citealt{Prochter,Vergani}) an excess of both weak and strong intervening absorbers have been found along the line of sight of GRBs compared to those of QSOs. These works consider an upper redshift limit of $z=2.3$ in order to compare the GRB statistics with the QSO
one, based on the SDSS spectra. If we consider this upper limit, the redshift path covered by the GRB100219A spectrum is $\Delta z=0.7$ and the detection of one strong \ion{Mg}{II} system within this $\Delta z$ is in line with an excess of the number density of these systems along GRB lines of sight. 

\cite{Sudilovsky07} performed also a statistical study on the GRB foreground CIV absorbers showing that their number is consistent with the one in QSO sightlines. The redshift path covered by our spectrum for the CIV absorbers is $\Delta z\sim1.15$, while the limit on the CIV column density is log N (cm$^{-2})\sim$14\, and 16 for b-parameters of 20 and 5\,km\,s$^{-1}$, respectively. \cite{Sudilovsky07} found 7 CIV systems in agreement with these limits within $\Delta z\sim2.25$, whereas we find none. We note that our  redshift range for CIV absorbers of $3.5\lesssim z\lesssim4.7$ is different from that of \cite{Sudilovsky07} ($1.5\lesssim z\lesssim3$) and no statistics are available for GRB CIV absorbers within this range. Thanks to the large spectral coverage of X-shooter, it will be possible to build a statistically significant and homogeneous sample of intervening absorbers and extend the studies to much higher redshifts.

\begin{table}[!hpt]
\caption{Column densities for the two velocity components in the intervening system at $z=2.181$.\label{tab:interv}}             
\label{tab:intervcomponents}      
\centering                          
\begin{tabular}{l | c c | c c  }        
\hline\hline                 
 &    \textbf{ I}  & \textbf{} & \textbf{ II}  & \textbf{}  \\    
   Ion, Transition    &$\Delta$v &log N &$\Delta$v&log N   \\ 
       \scriptsize {\hspace{6mm} \AA}		&\scriptsize (km/s)&\scriptsize (cm$^{-2}$)&\scriptsize (km/s)&\scriptsize (cm$^{-2}$)\\
\hline\hline  
\ion{Mg}{II} {\tiny 2796, 2803}  &  +52 & $13.10\pm0.08$ & --32 & $13.56\pm0.10$ \\           
\ion{Mg}{I} {\tiny 2852} & --- & --- &  --32 & $12.07\pm0.14$\\                 
\ion{Fe}{II} {\tiny 2600} & --- & --- &  --32  &  $13.13\pm0.12$\\        
\hline\hline
\end{tabular}
\end{table}

\section{GRB 100219A in the context of high redshift galaxy abundances}
\subsection{GRBs at $z>4$}\label{sect:z4grbs}
To date, 260 redshifts of GRBs have been determined (of these, 198 were detected by {\it Swift})\footnote{http://www.mpe.mpg.de/~jcg/grbgen.html and the GCN archive as of 11 June 2012}. This number includes secure photometric redshifts and redshifts determined using emission lines from the host galaxy, but not limits from optical/UV detections. Only 22 (8.5\%) of those 260 bursts are above $z=4$ (19 bursts or 9.6\% for {\it Swift)}, of which 4 were obtained by photometric redshift methods. 8 (3.1\% or 4.0\% for {\it Swift} only) are at $z>5$ of which 2 are photometric redshifts. The current records are $z=8.2$ for GRB 090423A (spectroscopic, \citealt{Salvaterra09,Tanvir09}) and $z=9.4$ for GRB 090429A (photometric, \citealt{Cucchiara}), which makes them one of the farthest objects detected in the Universe. 

\begin{figure}[!hpt]
\includegraphics[width=\columnwidth]{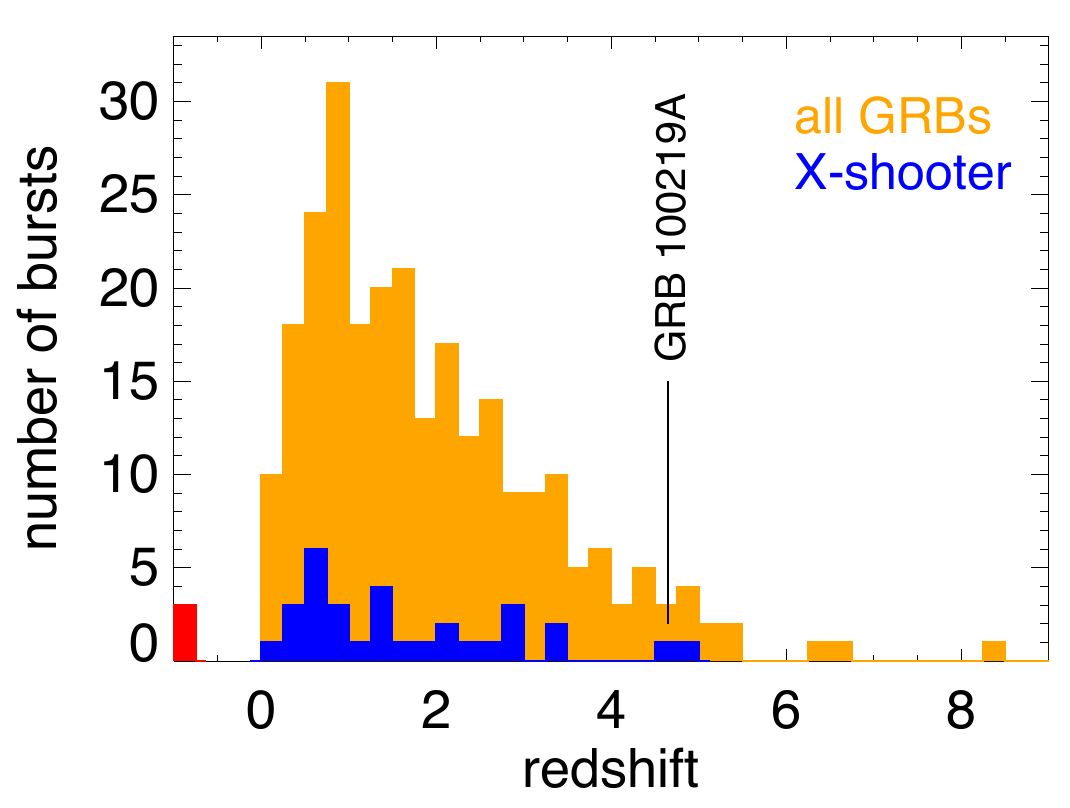}
\caption{Redshift distribution of all GRBs with secure redshifts (spectroscopic and photometric, orange) and the same for GRBs observed with X-shooter (blue histogram). X-shooter spectra where the redshift could not be determined are shown in red.
\label{redshiftdistr}}
\end{figure}

In the past, it had been estimated that a much larger fraction of GRBs might be observed at redshifts beyond $4-5$ \citep{Bromm02, Bromm05} and, taking into account the sensitivity of {\it Swift}, at least 10\% should be at z$>$5. It had been suggested that part of the high-redshift GRBs were missed due to the lack of NIR observations and high sensitivity NIR spectrographs to determine redshifts. However, \cite{Greinerdarkgrbs} detected afterglows with GROND for more than 90\% of the GRBs if observations were possible early enough, although they did not detect e.g. GRB 090429B at z$=$9.4 \citep{Cucchiara}. Their redshift distribution is similar to the overall distribution, leaving little room for a significant high-redshift population being missed by observations. X-shooter has to date determined 26 redshifts (2 or 8\% at z$>$4) and the redshift distribution for GRBs observed with X-shooter is similar to the distribution of all GRBs (see Fig. \ref{redshiftdistr}) \footnote{Since the first submission of this publication, X-shooter had detected a GRB with higher redshift than GRB 100219A, GRB 111008A at z$=$4.9898~\citep{WiersemaGCN}}. Observational constraints would also not explain the scarcity of GRBs between redshift $4-6$, which are still easily detectable with optical spectrographs. 

The reason for this low rate might be related to the capability of {\it Swift} (and other past and current $\gamma$-ray observatories) to detect high-$z$ GRBs. Time dilation stretches the emitted flux over a longer time interval, making rate triggers more difficult (in fact GRB 100219A was discovered in an image trigger). In addition, the peak emission of long, already intrinsically soft GRBs gets shifted to even lower bands, so that part of the flux falls below the BAT bands. Infact, \cite{Gorosabel04} concluded that the INTEGRAL satellite with it's different energy bands should detect more GRBs at high z than {\it Swfit}. Several authors have also suggested that without a luminosity evolution of GRBs it is unlikely that we would detect GRBs at that redshift at all (e.g., \citealt{Salvaterra07}), but determining the luminosity function of high redshift GRBs suffers from small number statistics. 

Another unknown factor is the evolution of the star-formation rate (SFR) at high redshifts. The peak of the SF is expected to be around $z\sim2$, and it is not yet known how the SF evolves at redshifts beyond $z\sim4$. Several attempts have been made to determine the cosmic SFR using GRBs (see, e.g., \citealt{Yuksel, Kistler09}) indicating a shallower decay towards higher redshift than derived by galaxy SF measurements. The redshift distribution of {\it Swift} GRBs at high redshifts can be well explained by a constant cosmic star-formation density from redshift 3.5 onwards \citep{Jakobsson06} \footnote{For a recent update see http://www.raunvis.hi.is/$\sim$pja/GRBsample.html}. {This could be explained if  GRBs preferentially select low-metallicity regions.} Which of the effects, an evolution of the luminosity function, a metallicity selection effect or a larger SFR at higher redshifts, is the one responsible for the observed evolution cannot be determined at present. 

In Table \ref{highzgrbs} we list the properties of all GRBs with $z>4$ with spectroscopic observations since the discovery of GRB afterglows. Of the 16 GRBs at $z>4$ with spectroscopic redshifts, 5 have measured metallicities, 3 more have derived limits. For GRBs with $4<z<5$, the majority has values or limits for the metallicity (7 out of 11), for the remaining 4 GRBs at least two seem to have good enough spectra to determine a metallicity but are not yet published (GRB 100302A, \citealt{ChornockGCN} and GRB 100513A, \citealt{CenkonewGCN}). The data of GRB 090516 which are available to our collaboration will be published in a forthcoming paper (A. de Ugarte Postigo et al., in prep.) and we list the metallicity determined from \ion{Si}{II} $\lambda$ 1808. This spectrum shows unusually strong lines, so that even the weak \ion{Si}{II} 1808 line is mildly saturated, so the real metallicity is probably $\sim0.3-0.4$\,dex higher, assuming typical values for the b-parameter around 20--30 km\,s$^{-1}$.

At $z>5$, determining metallicities gets increasingly difficult, only one metallicity has been reported for a GRB beyond $z=5$, GRB 050904 \citep{Kawai05}. However, they list metallicities derived from C, Si, O and S that differ by more than an order of magnitude and they adopted the value from S which is the highest from all 4 elements. They argue that this might be due to dust depletion, however, e.g., Si should track S rather close. \cite{Kann07} and \cite{Zafar10}, however, found no indication for dust depletion from the SED of the afterglow itself (but see \citealt{Stratta07,Stratta11}). 

We reinvestigate the spectra of GRB 050904 from \cite{Kawai05} which have also been studied in \cite{Zafar10} and find different EWs for S\,II $\lambda$ 1253 and Si\,II $\lambda$ 1260 (2.7 and 13.0 \AA{} observer frame respectively), while S\,II $\lambda$ 1259 is blended with Si\,II $\lambda$ 1260 and cannot be measured. All lines except S\,II listed in \cite{Kawai05} are completely saturated, also including another Si\,II line at $\lambda$1304 not listed in the paper. The saturation makes a curve-of-growth fit as mentioned in \cite{Kawai05} impossible, however, the column densities of Si\,II, O\,I and C\,II are definitely higher than listed in the paper. Taking the EW from S\,II $\lambda$ 1253 and assuming a (resonable) $b$-parameter of $\sim$ 35 km\,s$^{-1}$, we get a column density of around log N $\sim$ 15.3, 0.3 dex lower than listed in \cite{Kawai05}. \cite{Totani06} reinvestigated the DLA column density of GRB 050904 and find a slightly higher value of log N$_{HI}$=21.6. 

Together, this gives a revised metallicity of [M/H]$\sim$--1.6 and we give a conservative 3$\sigma$ error of 0.3 dex. This is also in line with the findings of \cite{Campana07} deriving a metallicity of [M/H]$>$--1.7 from X-ray spectra. Dust depletion is likely not playing a major role since the the metallicities from different elements are more similar than previously thought, although the definite values cannot be determined.

\subsection{Comparison between high redshift GRBs and QSO absorbers}
Observations of QSO absorbers at $z>4$ become difficult since current surveys only detect the most luminous ones. The intergalactic medium at those redshifts is almost opaque for Lyman$\alpha$ photons. This makes it difficult to detect and measure the column density of DLAs in the sightline, which in turn hampers the direct determination of the metallicity. The largest samples of QSO-DLAs and their properties \citep{Ledoux06, ProchaskaDLA, Prochaska05, Dessauges06} comprise a number of systems up to $z\sim4.5$. \cite{Becker11} determined properties of $5<z<6$ QSO absorbers but could only indirectly derive the metallicity of those systems.

QSO-DLAs show a decrease in metallicity with redshift \citep{Prochaska07, Dessauges06} with a large scatter at each redshift. There seems to be a ``metallicity floor'' at $\sim$ $\textnormal{[Z/H]}\sim-2.6$ explained by a rapid early enrichment due to star-formation. In Fig. \ref{fig:metal} we plot all GRB host metallicities reported in the literature together with metallicities from the large QSO-DLA sample of \cite{ProchaskaDLA} and fit a linear evolution to the data for each sample. As it has been pointed out in the past \citep{SavaglioNJP, Fynbo06}, GRB hosts have larger metallicities and less evolution with redshift compared to QSO-DLAs. At $z>4$ this becomes clearly evident as the mean GRB-DLA metallicity is about one order of magnitude larger than the mean of QSO-DLAs. Albeit a large scatter exists at all redshifts, the mean GRB-host metallicity seems to show a much shallower evolution compared to that of QSO-DLAs. 

\begin{figure}[!hpt]
\includegraphics[width=\columnwidth]{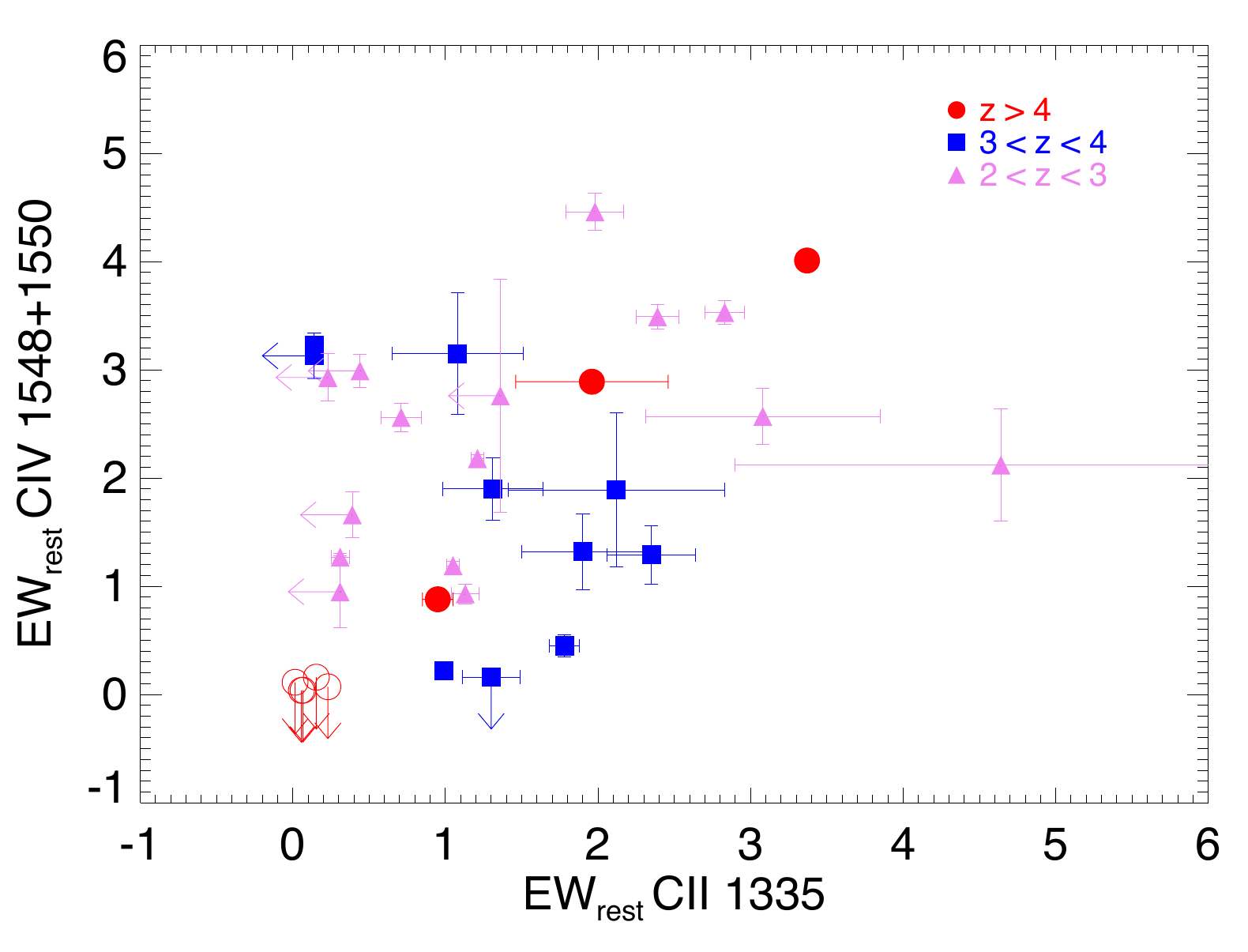}
\caption{Ionization of C in GRB-DLAs at 3 different redshift bins (filled symbols) and the comparison to measurements in $z=5-6$ QSO absorbers (empty circles) from \cite{Becker11}. GRB-EWs are taken from the low resolution {\it Swift} sample of \cite{Fynbo09} and its extension until Sept. 2009 published in A. de Ugarte Postigo et al. in prep. The CIV EWs for the GRB sample include both lines of the $\lambda\lambda$ 1548,1550 doublet, upper limits of the QSO sample are 2$\sigma$ limits on CIV $\lambda$1548.}Values for QSOs for the other two redshift bins are not shown for clarity and we refer to Fig. 13 in \cite{Becker11}. GRB-DLAs do not show any evident trend with redshift, neither in line strength nor in the ionization fraction and have higher line strengths than QSO absorbers at $z>$4. \label{fig:ionization}
\end{figure}

At $z>5$, the ionization fraction of QSO-absobers very low and \ion{Si}{IV} and \ion{C}{IV} are hardly ever detected \citep{Becker11}. However, this is consistent with the ionization fraction at lower redshifts due to a general decrease in column densities which makes the column densities of \ion{Si}{IV} and \ion{C}{IV} fall below the detection limit. GRBs with their generally higher metallicity should therefore show more \ion{Si}{IV} and \ion{C}{IV}, which is the case. GRBs do not exhibit any trend with redshift, neither in the overall abundance, nor is there any evident change in the ionization fraction and at high redshifts as shown in Fig. \ref{fig:ionization} for carbon. Special cases such as the highly ionized material around GRB 090426 (not shown in the plot) do not reflect this global picture since they likely probe very special environments \citep{Thoene10}. 

\begin{figure}[!hpt]
\includegraphics[width=\columnwidth]{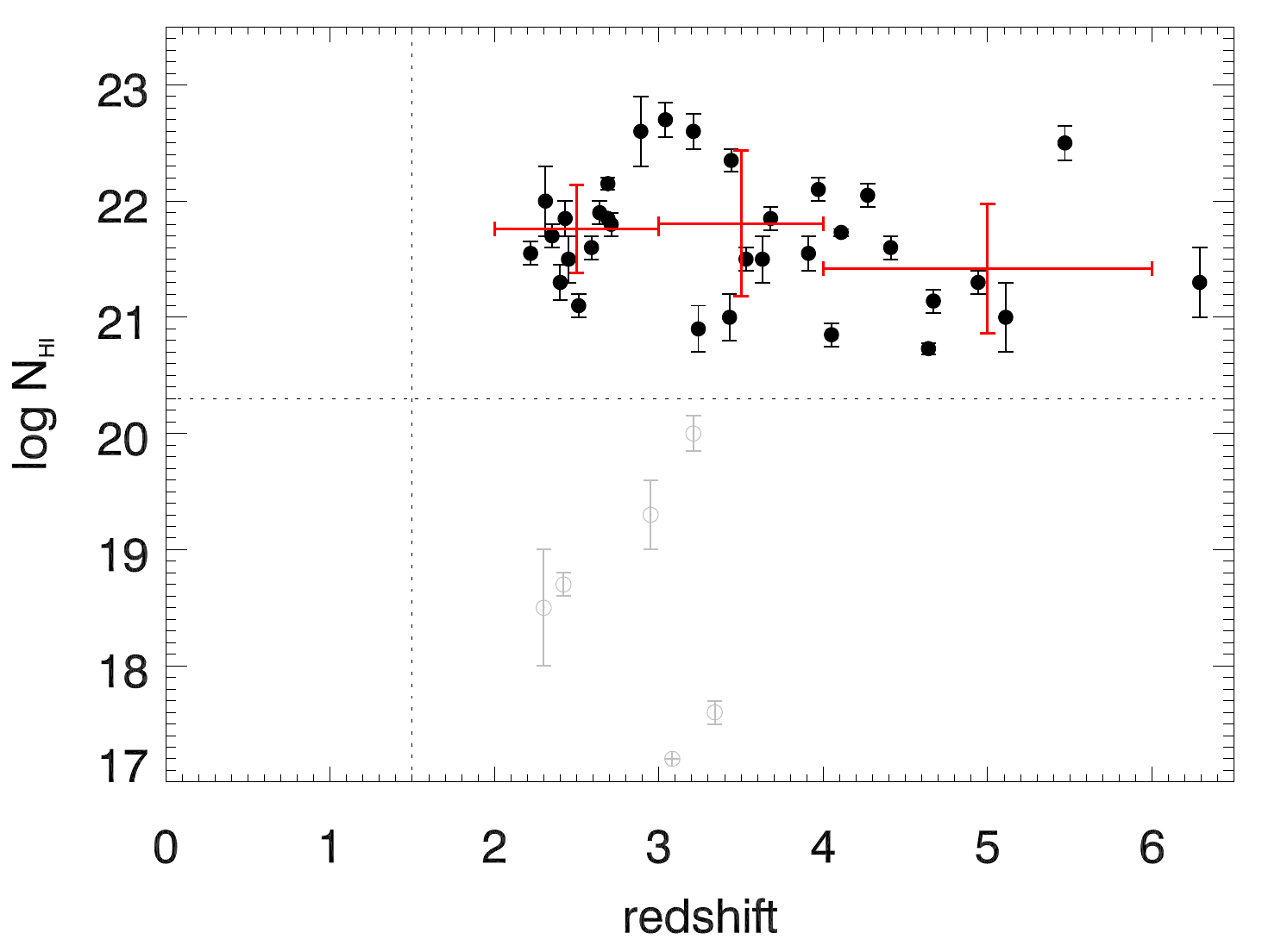}
\caption{Column density of {\it Swift} GRB-DLAs vs. redshift from the sample of \cite{Fynbo09} and later published values from the literature. The vertical dotted line indicates the detection limit of Ly $\alpha$ from the ground. Grey points are column densities, overplotted are the average column density and standard deviation in 3 redshift bins ($2<z<3$, $3<z<4$ and $z>4$). For the average column density we exclude sub-DLA systems (log N$_{HI}<20.3$, horizontal dotted line) since they might probe different environments; the values are plotted with empty circles. The GRB-DLA HI column density shows a large scatter at all redshifts but no trend towards lower column densities at higher redshifts as observed for QSO-DLAs. \label{fig:NH}}
\end{figure}

The distribution of GRB-DLA HI column densities have a higher mean value than those of QSO-DLAs and extend to much higher densities \citep[e.g.][]{Fynbo09}. It has been suggested that the cosmic neutral hydrogen density reaches a maximum around $z\sim3.5$ and decreases for lower redshifts \citep{Prochaska05}, since the gas is consumed or expelled from galaxies, while the evolution at $z>4$ is still inconclusive. The frequency distribution of DLAs and sub-DLAs shows little evolution but the slope might change slightly implying more sub-DLAs and less high column density systems at high redshifts \citep{Prochaska05, Guimaraes09}. In Fig. \ref{fig:NH} we plot log N$_{HI}$ for all GRBs for which this value is available from the literature and determine the mean and standard deviation at three redshift bins excluding sub-DLAs (log N$_{HI}<20.3$). The mean column densities are log N$_{HI}=21.76\pm0.38$ for $2<z<3$, $21.80\pm0.62$ for $3<z<4$ and $21.42\pm0.56$ for $z>4$, hence there is no evolution with redshift. Only a few GRBs with sub-DLA column densities have been found, all at $z<4$, but this is consistent with the number expected from the overall HI distribution. It seems GRB hosts have a high content of neutral gas irrespective of redshift which gives them a large reservoir for the formation of stars.

It is still debated whether QSO-DLAs (and GRB-DLAs) show an enhancement in $\alpha$ elements (O, Ne, Mg, Si, Ca), produced in massive stars, versus iron-peak elements (Fe, Ti, Zn), which primarily come from Type Ia SNe. An overabundance of $\alpha$ elements is expected for young star-forming regions and a corresponding correlation between metallicity and [$\alpha$/Fe] has been found for MW stars. Determining the  [$\alpha$/Fe] ratio is complicated by differential dust-depletion since Fe is more sensitive to dust depletion than e.g. Si. All studies of QSO-DLAs have found a statistically significant $\alpha$ enhancement (e.g., \citealt{Wolfe05, Dessauges06}). It has been claimed that, correcting for dust depletion, the enhancement disappears (e.g., \citealt{Pettini01} but see also \citealt{Savaglio00}). However, a mean positive ratio of [$\alpha$/Fe] was found even for very metal-poor (hence dust-free) systems \citep{Wolfe05} as well as for dust-free systems as determined by [Fe/Zn] (\citealt{Dessauges06}, Zn is supposed to be little affected by dust depletion). Extinction determined from the spectral continuum is generally low for GRB-DLA and QSO-DLAs with somewhat higher values for GRB-DLAs. However, dust depletion and extinction are not always tightly correlated, which could be caused by a ``patchy'' distribution of the dust along the line-of-sight. For GRB 100219A, we get a ratio [$\alpha$/Fe] of +0.39 (Si) and +0.76 (S). Si is only mildly depleted while S can be considered as little affected by depletion so the enhancement might be real. An enhancement has also been found for other GRBs \citep{SavaglioNJP} but the sample is still too small to make a definite conclusion or to investigate a possible redshift dependence.


\subsection{The mass-metallicity relation at high redshifts}
In the local Universe, galaxies follow a correlation between stellar mass content and metallicity and, with a less tight correlation, between ($B$-band) luminosity and metallicity (L-Z and M-Z relations). A corresponding mass-metallicity relation has been derived from the large SDSS sample by \cite{Tremonti04}, with metallicities derived from strong emission lines of the luminous gas in the galaxy. Deriving a reliable stellar mass content of galaxies usually requires broad-band SED fitting or at least a detection in the rest frame $K$ band, which makes it difficult to derive the masses at higher redshifts. Emission-line metallicity measurements at redshifts beyond $z\sim1$ require the use of high-sensitivity NIR spectrographs. In the past years, the local M-Z (or L-Z) relation has been extended to redshifts of $z\sim3.5$ (see \citealt{Mannucci09} and references therein). 

It has become clear that the M-Z relation at increasing redshift changes towards lower metallicities for a given stellar mass \citep{Kewley08,Savaglio05,Erb06,Mannucci09}. This effect is largely due to the enrichment with metals over time but the metal enrichment of galaxies is also a complex function of star-formation efficiency and feedback with the intergalactic medium. GRB hosts fall below the mass-metallicity relation for field (SDSS) galaxies \citep{Han, Kocevski} which had been taken as evidence for GRB hosts being biased towards lower metallicities, possibly because GRBs need low-metallicity environments \citep{Woosley05}. However, \cite{Mannucci10} established a connection between metallicity and star-formation rate showing that star-forming galaxies have on average lower metallicities, which can explain part of the discrepancy to field galaxies \citep{Kocevski}. Similar to the M-Z relation, the L-Z relation also shows an evolution with redshift, but the overall correlation is less tight. At low redshifts, the L-Z correlation is better constrained, probably due to a tight correlation between mass and luminosity. At higher redshift, the scatter in the mass-to-light ratio (M/L) between $B$-band magnitude and mass is rather large (see \citealt{Erb06}) and also steeper than in the local Universe, which makes the L-Z relation at $z>2$ rather unconstrained.

Metallicities of GRB hosts using emission lines have only been determined for a few hosts up to $z\sim2$, which will be significantly improved with X-shooter in the near future. At high redshifts, we have to rely on absorption-line metallicities, but we have to be aware that the metallicity derived from the luminous, hot gas of star-forming regions might not be the same as derived from the cold gas of the  galaxy ISM. We collected all reported host detections and upper limits for $z > 4$ GRBs from the literature (see Table \ref{highzgrbs}). Due to the lack of broad-band modeling, we only investigate the L-Z relation since the relation between $B$-band magnitude and stellar mass seems to change with redshift and is still uncertain at higher redshifts, therefore deriving a stellar mass would introduce additional errors. For GRB 050904, GRB 060223A, GRB 060510B and GRB 090423, rest-frame $B$-band magnitudes were given in \cite{Chary,Chary10} from {\it Spitzer} observations. For GRB 100219A, we take the $i'$-band detection from Sect. \ref{sec:photometry}. Magnitudes given in other bands were transformed to rest-frame B-band magnitudes using a rough k-correction of the form $k=-2.5 (1+\beta)\times  log_{10} (1+z\,(\lambda_{B}/\lambda_{init.}))$ with $\lambda_{init.}$ and $\lambda_{B}$ as the central wavelength of the original and the B-band filter respectively. We assume a powerlaw slope for the host SED of the form $F_\nu \propto \nu^ {\beta}$ with $\beta = 0.5$.

\begin{figure}[!hpt]
\includegraphics[width=\columnwidth]{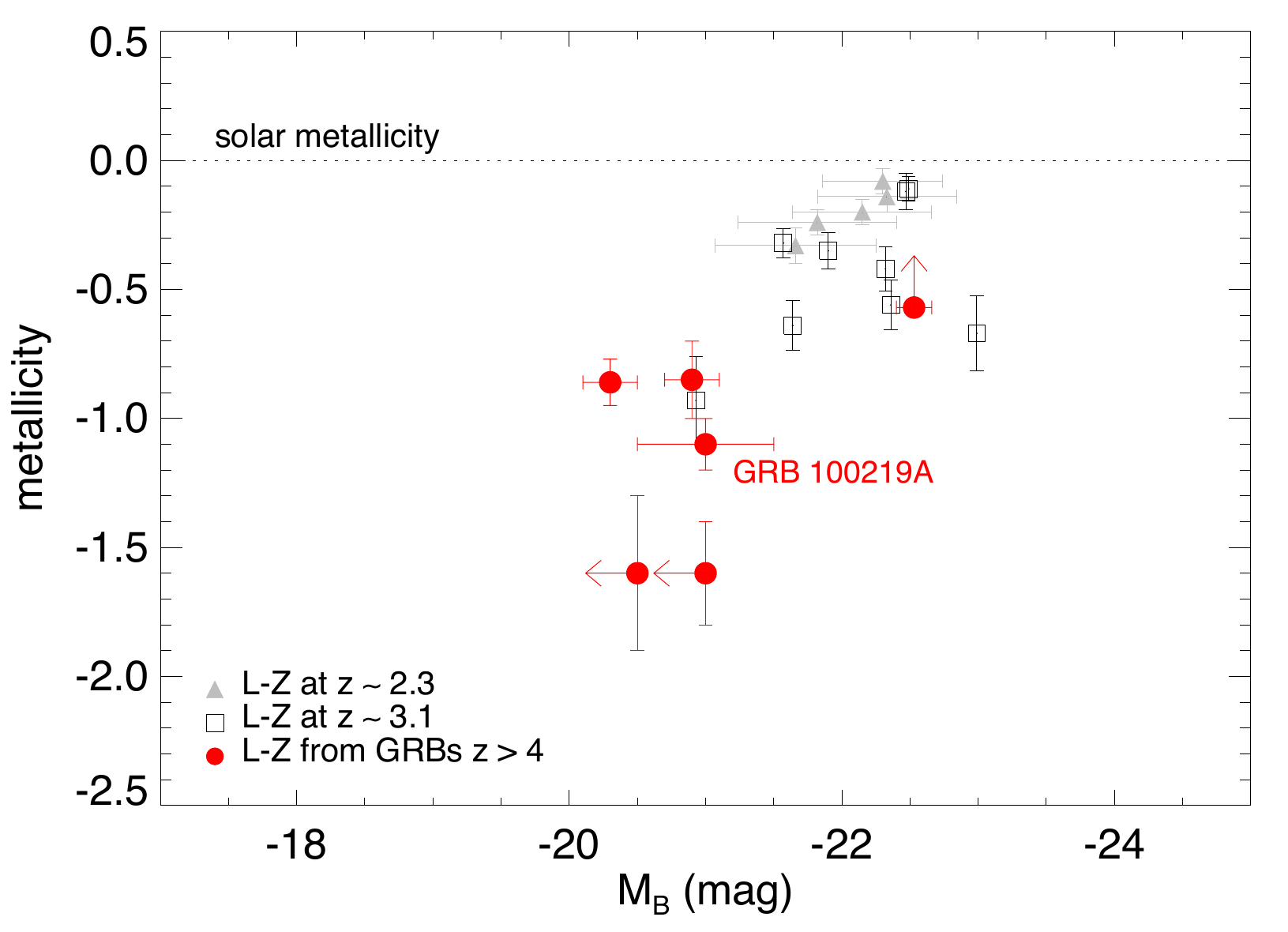}
\caption{Luminosity-metallicity relation of $z > 4$ GRBs compared to field galaxy samples at slightly lower redshifts. Galaxy data are from \cite{Erb06} for $z\sim2.3$ and from \cite{Mannucci09} for $z\sim3.1$. \label{fig:MZ}}
\end{figure}

Fig. \ref{fig:MZ} shows the result for GRB hosts with $z > 4$ compared to two other samples for field galaxies of $z\sim2.3$ \citep{Erb06} and $z\sim3.1$ \citep{Mannucci09}. GRB 050904 at $z = 6.29$ with our corrected metallicity and GRB 060223A clearly lie below the galaxies at $z\sim3.1$ while the other four hosts, including GRB 100219A, are consistent with the relation at lower redshifts. The number of high-redshift GRBs is still too low to make any conclusive statement on the L-Z relation of GRB hosts. 

\cite{Salvaterra11} studied the redshift evolution of the mass-metallicity relation from redshift 5 to 10 using galaxy evolution simulations with the GADGET code \citep{Springel05}. They found considerable enrichment even of the smallest galaxies of up to 1/10 already at redshifts $7-8$ and probably even higher. The metallicity of $\sim0.1$ Z$_\odot$ for GRB 100219A is fully consistent with chemical evolution models. Comparing the L-Z relation at $z=5$ from the simulations of \cite{Salvaterra11}, our GRB hosts at $z=4-5$ would still fall somewhat below the relation, confirming the trend seen at lower redshifts. Further investigations of the relation with new observations, in particular at z$>$\,5 would be highly desirable to study the early chemical enrichment of galaxies and their star-formation history and ISM feedback. 

A direct determination of the L-Z/M-Z relation of QSO absorbers is difficult to obtain, however, several authors have confirmed a correlation of the velocity spread of, e.g., \ion{Si}{II} $\lambda$ 1526 and the metallicity \citep{Ledoux06, Prochaskavelfield}. This correlation seems to decrease with redshift, mostly due to the change in metallicity. The velocity spread is probably related to the dark-matter halo mass of the absorbers \citep{Ledoux06}, hence the velocity-metallicity relation can be seen as a mass-metallicity relation and the observed trend fits to the M-Z/L-Z relations found in emission. A tight correlation has also been found for the metallicity vs. the EW of \ion{Si}{II} $\lambda$ 1526 of QSO-DLAs such that it can be used as a proxy for metallicity if no other measurement is available \citep{Prochaskavelfield} and this correlation might also be present for GRB-DLAs, albeit with a different slope and offset. The reason for this correlation might be the underlying velocity-metallicity relation and the suggestion that the EW of saturated absorption lines traces the kinematics of the gas well. The EW of \ion{Si}{II} $\lambda$ 1526 of GRB-DLAs does not evolve much with redshift \citep{Fynbo09}, while it decreases for QSO-DLAs. On the other hand, the velocity spread of, e.g., \ion{Si}{II} $\lambda$ 1526 in QSO-DLAs does not show large changes at high redshifts \citep{Prochaska05, Becker11}. The identification of the galaxy counterparts of QSO-DLAs is difficult due to their faintness and the proximity of the QSO, therefore, a direct mass- or luminosity-metallicity relation for the galaxy counterparts of QSO-DLAs has not been determined yet. 

\begin{table*}[!hpt]
\centering
\caption{Properties of all z$>$4 GRBs. Those with only a photometric redshift, GRB 050814, \citep{Jakobsson06}, GRB 071025 \citep{Perley10}, GRB 080916C \citep{Greiner09c}, GRB 080825B  \citep{Kruehler11} and GRB 090429A \citep{Cucchiara}, are not included. Metallicities in bold are the ones adopted throughout the paper. 
References for values from the literature are: [1] \cite{Andersen00}, [2] \cite{Berger05}, [3] \cite{Kawai05}, [4] \cite{Totani06}, [5] \cite{Berger07} [6] \cite{Chary}, [7] \cite{Thoene08}, [8] \cite{Price}, [9] \cite{RuizVelasco}, [10] \cite{Greiner09b}, [11] \cite{DAvanzo10}, [12] \cite{Greiner09a}, [13] \cite{Patel10} [14] \cite{Tanvir09}, [15] \cite{Salvaterra09}, [16] \cite{Chary10}, [17] Gorosabel et al. in prep., [18] \cite{ChornockGCN}, [19] \cite{CenkoGCN}, [20] \cite{WiersemaGCN}, [21] J. Fynbo, priv. comm. \label{highzgrbs}}
\begin{tabular}{lllllll}		\hline
GRB & z & log N$_\mathrm{HI}$&[Z/H] & host M$_\mathrm{B}$& ref&comment\\	
	&	&(cm$^{-2}$)&	&(AB mag)& \\		
\hline\hline									
000131&4.50&---&---&---&[1]&only Ly$\alpha$ detected\\
050505&4.27&$22.05\pm0.1$&{\bf --1.2} (S), --1.6 (Si), --2.0 (Fe)&---&[2]&---\\
050904&6.29&$21.6$&--1.0 (S), --2.4 (C), --2.3 (O), --2.6 (Si)&$>-20.5$ &{\bf [3,4,5]}& \\
		&	&	&{\bf --1.6\,$\pm$\,0.3}&&this work& revised metallicity\\
060223A&4.41&$21.6\pm0.1$ & $<-1.45$ (Si), $>-1.8$ (Si)&$>-21.0$&[6]&\\
		&	&	&{\bf --1.6\,$\pm$\,0.2}&&\\
060206&4.04&$20.85\pm0.1$&{\bf --0.86} (S), --1.08 (Si), --1.16 (C)&$-20.3\pm0.2$ &[7]&\\
060510B&4.94&$21.3\pm0.1$&{\bf --0.85} (S), $>-0.8$ (Fe)&$-20.9\pm0.2$&[6,8]&\\
060522&5.11&$21.0\pm0.3$&---&$>-26.4$&[6]&only L$\alpha$  detected\\
060927&5.47&$\sim22.5$&---&---&[9]&low S/N spectrum\\
080129&4.35&---&---&---&[10]& low S/N spectrum\\
090205&4.64&$20.73\pm0.05$&{\bf $>$ --0.57} (S)&$-22.5\pm0.1$ &[11]&\\
080913&6.69&19.84&---&---&[12,13]&\\
090423&8.2&---&---&$>-19.9$&[14,15,16]&\\
090516&4.11&$21.73\pm0.02$&{\bf $>$ --1.2} (Si) &---&[17]&\\
100302A&4.81&---&---&---&[18]&\\
100219A&4.67&$21.14\pm0.04$&{\bf --1.1} (S), --1.4 (Si), --1.9 (Fe), --1.0 (O)&$>-22.9$&this work&\\
100513A&4.77&---&---&---&[19]&\\
111008A&4.98&$\sim$22.4&---&---&[20,21]&\\
\hline	
 \hline
\end{tabular}
\end{table*}

\section{Conclusions}
In this paper we presented X-shooter spectra of GRB 100219A at $z=4.667$ which is the highest redshift at which a high-resolution spectrum has been available until now, allowing a detailed analysis of the abundances in a high-redshift (star-forming) galaxy. The metallicity as determined from S is moderately high with $\textnormal{[M/H]}=-1.0$ or 0.1 Z$_\odot$, similar to the value found for other GRBs between redshift $3-4$, but 10 times higher than the average metallicity of galaxies found in the sightlines of QSOs at this redshift. There is little evidence for extinction from the afterglow SED and the relative abundances of the detected metal absorption species indicate only a mild dust depletion. This suggests that the $\alpha$-element enhancement of [$\alpha$/Fe]$=$0.3--0.7 we find from Si and S is likely real. The kinematics of low-, high-ionization and fine-structure lines is rather complicated, probably showing an early galaxy in the process of formation or merging with another galaxy. We also detect one intervening system at $z=2.181$, consistent with the detection rate of intervening absorbers in GRB sightlines.

Studying galaxies and their abundances in the high-redshift Universe is of great importance to our understanding of the cosmic chemical evolution, in particular at redshifts where galaxies were still in the process of formation. Over the last years it has become evident that galaxies probed by QSOs and GRBs show somewhat different properties. While GRB hosts have a very shallow metallicity evolution, QSO-DLAs have on average a factor of 10 times less in metallicity at $z\sim4$ and show a steep increase up to $z\sim2$, where the metallicities of both samples become similar. GRBs also show no evolution in ionization rate (though this is likely also the case for QSO-DLAs) and in the HI average column density and distribution. Probably, GRBs select similar types of galaxies across the history of the Universe, highly star-forming and metal-poor, while QSO-DLAs probe the average galaxy population.

It has been suggested that both QSO- and GRB-DLAs might come from the same population of galaxies and the metallicity distribution can be explained by a sightline effect (GRBs probe the dense regions of the galaxy while QSO-DLAs probe the more metal-poor outskirts) combined with a slightly higher average mass for GRB-DLA galaxies. This seems likely since GRB-DLAs also have much stronger lines, hence probing denser regions, and a higher velocity width than QSO-DLAs. It is still unclear into what type of galaxies in the present Universe high-redshift QSO-DLAs develop into, though they are likely not dwarf galaxies \citep{Prochaska05}. GRB hosts at low redshifts are a diverse mix of star-forming galaxies, from blue compact dwarfs to spiral galaxies. 

The picture is somewhat inverted for the mass-metallicity relation of GRB hosts compared to field galaxies. At low redshifts, GRBs select galaxies more metal-poor for their mass than average, even after accounting for a possible bias from star-forming galaxies (that have shown to be on average more metal-poor). The comparison at higher redshifts is still unclear but the difference might be less pronounced than in the local Universe. It will be an important task to develop a common picture for the galaxy populations probed by different methods in order to get the global picture of galactic chemical evolution. 

The evolution beyond redshift $\sim5$ is still largely unknown since the current detection rate of such high-redshift GRBs is still low. At redshifts approaching the reionization epoch and the formation of the first stars and galaxies, the determination of metallicities and relative abundances are crucial to test the models for high-redshift star-formation and the formation and evolution of galaxies.

\acknowledgements
CT acknowledges the hospitality of DARK where part of this work was done and thanks A. Kann and D. Malesani for carefully reading the  final manuscript. This work is based on observations made with X-shooter under the guaranteed time program 084.A-026 (PI: Fynbo), GROND at the MPI/ESO 2.2~m telescope and with telescopes at the European Southern Observatory at LaSilla/Paranal, Chile under program 084.D-0764, PI: Greiner. This work made use of data supplied by the UK Swift Science Data Centre at the University of Leicester.  The ``Dark Cosmology Centre'' is funded by the Danish National Research Foundation. Part of the funding for GROND (both hardware as well as personnel) was generously granted from the Leibniz-Prize to Prof. G. Hasinger (DFG grant HA 1850/28-1). JPUF and BMJ acknowledge support from the ERC-StG grant EGGS-278202. TK acknowledges support by the DFG cluster of excellence 'Origin and Structure of the Universe' and support by the European Commission under the Marie Curie Intra-European Fellowship Programme in FP7. SK and ANG acknowledge support by DFG grant Kl 766/16-1. CT and J. Gorosabel acknowledges support by the Spanish Ministry of Science and Innovation under project grants AYA2009-14000-C03-01, AYA2010-21887-C04-01 (``Estallidos'') and AYA2011-24780/ESP (including Feder funds). SCa, SCo, GT, SP acknowledge support from ASI grant I/011/07/0.

\end{document}